\documentclass[epj]{svjour}
\usepackage{graphicx}
\usepackage{epsfig}
\usepackage{epstopdf}
\usepackage[english]{babel}
\usepackage{color}
\usepackage{amsmath}
\usepackage{amssymb}
\usepackage{amsfonts}
\usepackage{bm}
\usepackage{bbm}
\usepackage{cite}

\usepackage{xcolor}
\usepackage{hyperref}
\hypersetup{
    colorlinks,
    linkcolor={red!80!black},
    citecolor={blue!50!black},
    urlcolor={blue!80!black}
}

\def\red#1{#1} %

\newcommand{\ket}[1]{|#1\rangle}			
\newcommand{\bra}[1]{\langle#1|}			
\renewcommand{\d}{\mathrm{d}}				
\newcommand{\I}{\mathbbm{1}}				

\newcommand{\sigmas}{{\{\sigma_i\}}}			
\newcommand{\Sz}{S^{z}}					
\newcommand{\Sx}{S^{x}}					
\newcommand{\Tr}{\mathrm{Tr}}				
\newcommand{\bracket}[2]{\langle#1|#2\rangle}		

\begin{document}
\title{Effective One-Dimensional Models from Matrix Product States}
\subtitle{}
\author{Frederik Keim\thanks{\email{frederik.keim@tu-dortmund.de}} \and G\"otz S. Uhrig\thanks{\email{goetz.uhrig@tu-dortmund.de}}
}                     
\institute{Lehrstuhl f\"{u}r Theoretische Physik I, TU Dortmund, Otto-Hahn Stra\ss{}e 4, 44221 Dortmund, Germany}
\date{Received: 9 March 2015 / Recieved in final form 28 April 2015\\
The final publication is available at Springer via
\href{http://dx.doi.org/10.1140/epjb/e2015-60188-0}{http://dx.doi.org/10.1140/epjb/e2015-60188-0}}
%
\abstract{
In this paper we present a method for deriving effective one-dimensional models based on the matrix product state formalism.
It exploits translational invariance to work directly in the thermodynamic limit.
We show, how a representation of the creation operator of single quasi-particles in
both real and momentum space can be extracted from the dispersion calculation.
The method is tested for the analytically solvable Ising model in a transverse magnetic
field.
Properties of the matrix product representation of the creation operator are discussed and validated by calculating the one-particle contribution to the spectral weight.
Results are also given for the ground state energy and the dispersion.
\PACS{
	{02.70.-c}{Computational techniques; simulations}
	\and
	{75.10.P }{Spin chain models}
	\and
	{05.10.Cc}{Renormalization group methods}
     } 
} 

\maketitle

\section{Introduction}
\label{sec:intro}

Strongly correlated quantum systems remain a great challenge in condensed matter theory.
In many cases, we have to rely on numerical tools to make predictions from a microscopic model that can be compared to experiment.
Unfortunately the exponential growth of the Hilbert space dimension severely limits the size of systems that can be treated exactly.

However, to gain understanding of the physics in quantum systems, often the properties of the ground state and a few excited states go a long way.
Therefore, a variety of tools have been developed to separate the low energy part of the Hilbert space from the rest.
In one dimension, this is most notably the density matrix renormalization group (DMRG) \cite{white92}, which is a variational ansatz.
There are various extensions of the original DMRG, many of which work within the framework of matrix product states (MPS) \cite{schol11} or extensions thereof
\cite{vidal03,shi06}.

Another approach that extends more readily to higher dimensions is the semi-numerical method of continuous unitary transformations (CUTs) \cite{knett00a,knett03a,kehre06,yang11a,krull12}.
The idea behind these CUTs is to partially diagonalize the Hamiltonian and derive an effective model (explained below) in second quantization for the low-energy sector of the Hilbert space.
This effective model can then be used to calculate physical properties of the system.
CUTs have been applied successfully in a variety of cases
\cite{moeck08,dusue10,fause13a}.

In our view, it is a promising long-term goal to establish effective models
for strongly correlated systems in terms of the elementary excitations with the
ground state being the vacuum. Assuming translational invariance for infinitely large
systems, i.e., systems in the thermodynamic limit, the momentum space notation is convenient due to momentum conservation and mutual orthogonality of momentum eigenstates with different momenta.
The goal of our approach is to map a microscopic lattice model such as the transverse field Ising model (TFIM), see Eq.\ \eqref{eq.H_TFIM}, to an effective Hamiltonian for the low-energy physics in the quasi-particle picture
\begin{subequations}
\label{eq.H_eff}
\begin{align}
  \mathcal{H}^\mathrm{eff} &= E_0 + \sum_{q_1} \omega_{q_1} a_{q_1}^\dagger a_{q_1} \label{eq.H_eff_1qp} \\
  & + \frac{1}{\sqrt{L}} \sum_{q_1, q_2, q_3} \left[ D^{q_1}_{q_2,q_3} a^\dagger_{q_2} a^\dagger_{q_3} a_{q_1} + \mathrm{h.c.} \right] \, \delta_{q_2+q_3,q_1}
	\label{eq.H_decay} \\
  & + \frac{1}{L} \sum_{\begin{matrix} q_1, q_2,\\ q_3, q_4 \end{matrix}} V^{q_1, q_2}_{q_3, q_4} \, [a_{q_1}^\dagger a_{q_3}^\dagger a_{q_2} a_{q_4}] \,
	\delta_{q_1+q_2,q_3+q_4}
	\label{eq.H_int} \\
  & + [\mathrm{ higher\ terms }] \nonumber \ .
\end{align}
\end{subequations}
In $\mathcal{H}^\mathrm{eff}$ the second quantization operator $a_{q_{i}}^\dagger$ creates a quasi-particle with momentum $q_{i}$ and $a_{q_{i}}$ annihilates one.
The ground state energy is labeled $E_0$ and the dispersion relation $\omega_q$.
The $D^{q_1}_{q_2,q_3}$ and $V^{q_1, q_2}_{q_3, q_4}$ are the matrix elements for quasi-particle decay and two-particle interaction, respectively.
Momentum conservation is included through the Kronecker $\delta$ symbols.
At this point we do not include further algebraic properties, since these depend on the model under consideration.

By the effective model \eqref{eq.H_eff},
the fundamental physical properties of complicated microscopic
models can be reduced to a level which makes further quantitative studies
possible. The above mentioned continuous unitary transformations (CUTs) represent a
systematic tool which successfully achieves this aim \cite{knett03a,fisch10a}.
They work particularly well if the chosen starting point, the so-called
reference state, is already close to the true ground state of the system.
But if this is not the case or if it is even not possible to find
a technically tractable starting point with the relevant symmetries
the CUTs cannot be applied efficiently. In such situations a completely
numerical approach is appealing because it can tackle a problem at hand
in an unbiased fashion.

This is where the MPS representation comes into play.
It is extremely efficient in finding ground states and certain excited states.
We will show how one can use MPS to numerically define and derive
local creation and annihilation operators $a^{(\dag)}_i$ at site $i$.
Thereby, we provide the first steps towards representations such as
\eqref{eq.H_eff} derived by numerical variational means.

The key to efficient handling of translationally invariant systems are transfer matrices.
Their usefulness in handling MPS representations of infinte systems has been known for a while \cite{fanne92}.
Impressive progress has been achiev{-}ed in describing elementary excitations
within the MPS framework \red{\cite{haege12,haege13a,Haegeman2013b,Haegeman2014}}.
Haegeman et al.\ presented a momentum space ansatz \cite{haege12,Haegeman2013b} that yields very accurate results for the dispersion relation and is
related to the calculation presented in this paper.
They also proved by means of the Lieb-Robinson bounds \cite{lieb72} that an excited momentum eigenstate
of a lattice Hamiltonian can be exponentially well approximated by acting on the ground state with the
momentum superposition of a local operator with finite support \cite{haege13a} if the system displays an energy gap.

In the present paper we show how such an operator can be constructed using eigenvectors
of an eigenvalue problem arising in the dispersion calculation.
We demonstrate, that this operator is a representation of the local creation operator
$a^\dagger$ by using it to compute the one-particle contribution to the spectral weight. \red{To this end, we think that a brief presentation of the concept of matrix product states is required in order to explain technical details
and the employed notation.}

We test the method on the TFIM which is analytically solvable \cite{pfeut70} and well understood.
Inspite of its simplicity, it shows a variety of interesting features such as a quantum phase transition, ground state degeneracy and two different types of elementary excitations.

The paper is structured as follows: In Section \ref{sec:model} the model and its exact solution are recalled.
In Section \ref{sec:MPS} a short introduction to matrix product states is
given\red{, which may be skipped by readers familiar with this concept.}
Section \ref{sec:GSE} shows the results for the ground state energy and the dispersion followed by the derivation of the effective model in Section \ref{sec:eff_model}.
In Section \ref{sec:spectral_weight} we compute the static one-particle spectral weight as an application of the effective model and finally conclude the paper in Section
\ref{sec:outlook}.

\section{Model and exact results}
\label{sec:model}

The TFIM \cite{genne63} is a common toy model for studying quantum magnets.
In the one-dimensional ferromagnetic case considered here, it is given by the Hamiltonian
\begin{equation}
  \mathcal{H} = - \Gamma \sum_i \Sz_i - J \sum_i \Sx_i \Sx_{i+1}, \quad \Gamma,\ J > 0 \label{eq.H_TFIM}
\end{equation}
where $\Sz$ and $\Sx$ are components of the standard spin-$\frac{1}{2}$ operators, $\Gamma$ is the external field and $J$ the coupling strength.
We define the dimensionless parameter
\begin{equation}
  \lambda = \frac{J}{2 \Gamma} \label{eq.def_lambda}
\end{equation}
that controls the system's behavior and in terms of which the Hamiltonian reads
\begin{equation}
  \mathcal{H}_\lambda := - \sum_{i} \Sz_i - 2\lambda \sum_{i} \Sx_i \Sx_{i+1} \ .
\end{equation}
At the point $\lambda = 1$ the coupling to the external field is of the same strength as the nearest-neighbor coupling, giving rise to a quantum critical point.

The phase with $\lambda > 1$ is called the Ising regime or ordered phase since the Ising interaction is dominant.
The ground state is twofold degenerate since its long-range order spontaneously breaks the Hamiltonian's $\mathbb{Z}_2$ symmetry $\Sx \to -\Sx$.
In 1D, the elementary excitations are (dressed) domain walls between regions of the two different ground states.

Conversely, for $\lambda < 1$ the external field dominates the behavior.
This phase is called the strong-field regime or disordered phase.
Due to the strong external field, its ground state is unique and its excitations are
(dressed) spin flips.
The model is analytically solvable by a sequence of Jordan-Wigner, Fourier and Bogoliubov transformations \cite{parki10} as shown by Pfeuty
\cite{pfeut70} \red{and has been studied extensively, see for instance
Refs.\ \cite{mccoy83a,mccoy83b,mulle85,perk09}.
Here we will recall the important known facts.}

The ground state energy per lattice site is given by the elliptic integral
\begin{align}
  \frac{E_0}{L} &= - \frac{\Gamma (1+\lambda)}{\pi} \int_0^{\frac{\pi}{2}} \sqrt{ 1 - \frac{4\lambda \, \sin^2q}{(1+\lambda)^2}} \, \d q
	\label{eq.E0_TFIM}
\end{align}
which displays a non-analyticality at $\lambda = 1$ due to the singular derivative of the square root function.

The one-particle dispersion relation $\omega_q$ reads
\begin{align}
  \omega_q &= \Gamma \sqrt{1 + \lambda^2 - 2\lambda \cos q}
	\label{eq.Dispersion_TFIM}
\end{align}
with the lattice constant set to unity.
From this the excitation gap $\Delta$ is read off to be
\begin{align}
  \Delta = \min_q \omega_q = \Gamma | 1 - \lambda | \label{eq.Gap_TFIM}
\end{align}
which vanishes at $\lambda = 1$.

A quantity of great interest in connection with elementary magnetic excitations is the ground state spin-spin correlation function defined by
\begin{align}
  G^{\alpha\beta}_j := \langle S^\alpha_0 S^\beta_j \rangle = \bra{\psi_0} S^\alpha_0 S^\beta_j \ket{\psi_0} \label{eq.def_Gj} \ .
\end{align}
where $\alpha, \beta \ \in \ \{x, y, z, +, -\}$ and  $\ket{\psi_0}$ denotes the ground state.
For translationally invariant gapped systems with local Hamiltonians in one dimension the ground state correlation function Eq.\  \eqref{eq.def_Gj} is known to show exponential behavior \cite{hasti06}
\begin{align}
  G_j \propto \exp\left( - \frac{|r_j|}{\xi} \right) \label{eq.Gj_asymptotics}
\end{align}
where $\xi$ is the correlation length.
For the 1D TFIM it can be calculated analytically \cite{okuni01} to be
\begin{align}
  \xi = \frac{1}{|\ln\lambda|}
	\label{eq.xi_TFIM} \ .
\end{align}
The standard approximation $\xi \approx \frac{v}{\Delta}$ where $v$ is obtained by fitting $\omega_q \approx \sqrt{\Delta^2 + (2v \sin\left(\frac{q}{2}\right))^2}$ to the minimum and maximum of the dispersion is in very good agreement with Eq.\
\eqref{eq.xi_TFIM} for $\lambda \gtrsim 0.2$.

Another important quantity in relating theoretical results to experiment is the dynamical structure factor (DSF) \cite{marsh71}
\begin{align}
  S^{\alpha\beta}(\omega,q) = \frac{1}{2\pi L} \sum_{i,j} \int_{-\infty}^{\infty} \d t \, e^{i[\omega t + q(r_i - r_j)]}
  \left\langle S^{\alpha}_j(t) S^\beta_i(0) \right\rangle
	\label{eq.def_DSF}
\end{align}
which describes the intensity in neutron scattering experiments.
We consider only zero temperature where the angular brackets denote the ground state expectation value.\\
Integrating Eq.\  \eqref{eq.def_DSF} over frequency yields the static structure factor
\begin{align}
  S^{\alpha\beta}(q) = \frac{1}{L} \sum_{i,j} e^{iq(r_i - r_j)} \langle S^\alpha_j S^\beta_i \rangle \label{eq.def_SSF}
\end{align}
which is the Fourier transform of the ground state spin-spin correlation function Eq.\  \eqref{eq.def_Gj}.

If for given momentum $q$ the energy levels are well separated, i.e. the spectral function is a sequence of Dirac-$\delta$-spikes, Eq.\  \eqref{eq.def_DSF} can be written in the spectral form
\begin{align}
  S^{\alpha\beta}(\omega,q) = \sum_\Lambda \delta(\omega - E_\Lambda) S^{\alpha\beta}_{\Lambda}(q) \ .
	\label{eq.def_SLambda}
\end{align}
The spectral weights $S^{\alpha\beta}_{\Lambda}(q)$ are given by projecting
Eq.\ \eqref{eq.def_SSF} onto the states with energy $E_\Lambda$.
Note that the energy $E_\Lambda$ is understood to be defined \emph{relative}
to the ground state energy.

Except at criticality, the TFIM has a well defined one-particle energy $\omega_q$.
Thus evaluating Eq.\  \eqref{eq.def_SLambda} at $E_\Lambda = \omega_q$ is valid and yields the one-particle spectral weights
\begin{subequations}
\begin{align}
  S^{\alpha\beta}_{1\mathrm{p}}(q) &= \frac{1}{L} \sum_{j,j'} \bra{\psi_0} S_i^\alpha \ket{\phi_q}\bra{\phi_q} S_j^\beta \ket{\psi_0} e^{iq(j'-j)}
	\label{eq.def_Sab} \\
  &= \bra{\psi_0} S_q^{\alpha\dagger} a_q^\dagger \ket{\psi_0} \bra{\psi_0} a_q S_q^\beta \ket{\psi_0}
	\label{eq.Sxxq}
\end{align}
\end{subequations}
where $\ket{\phi_q}$ is a one-particle state with energy $\omega_q$.
\red{Hamer et al.\ have given an analytic formula for the spectral weight in the
$xx$-channel $S^{xx}_\mathrm{1p}(q)$ in the disordered phase.
The expression has been conjectured by them
from high order series expansion \cite{hamer06b}.
In fact, its Fourier transform
had been derived exactly by Vaidya and Tracy \cite{vaidya78}.}
It reads
\begin{align}
  S^{xx}_\mathrm{1p}(q) = \frac{(1-\lambda^2)^\frac{1}{4}}{\omega(q,\lambda)} \ , \quad \lambda < 1
	\label{eq.Sxx_Hamer} \ .
\end{align}
Note, that there is no single-particle contribution in the $xx$-channel for
$\lambda > 1$ \cite{vaidya78}.

\section{Matrix product states}
\label{sec:MPS}

\subsection{Definition}

The formalism of matrix product states (MPS) has been introduced in various contexts \cite{baxte68,fanne89,fanne92}. It is a way of denoting quantum mechanical states that is particularly convenient for variational calculations.
It is also closely related to the DMRG method
\cite{ostlu95,romme97,schol11}. This section gives a brief introduction to the concept.
Since we are interested only in translationally invariant chain models, we will restrict ourselves to those here.
For a more detailed overview, we refer the reader to Ref.\ \cite{schol11}.

Consider a state $\ket\psi$ of a system with $L$ sites where $\sigma_i$ defines
the local state at each site $i$
\begin{equation}
  \ket{\psi} = \sum_{\sigma_1,\ldots,\sigma_L} c_{\sigma_1,\ldots,\sigma_L}
	\ket{\sigma_1,\ldots,\sigma_L}
  = \sum_{\{\sigma_i\}} c_{\{\sigma_i\}} \ket{\sigmas}
\end{equation}
where the $\ket{\sigma_1,\ldots,\sigma_L}$ represent an ortho normal basis set.
For simplicity, we assume that all $\sigma_i$ have the same local Hilbert space dimension $d$.
The expansion coefficients $c_{\sigma_1,\ldots,\sigma_L}$ can be interpreted as elements of a matrix $\Psi^{[1]}_{(\sigma_1),(\sigma_2,\ldots,\sigma_L)}$ of dimension $d \times d^{L-1}$.
This can be written as
\begin{align}
   \Psi^{[1]} = U^{[1]} S^{[1]} V^{\dagger\,[1]}
	\label{eq.SVD}
\end{align}
by means of the singular value decomposition (SVD).
In Eq.\  \eqref{eq.SVD} $U^{[1]}$ is a $d\times d$ unitary matrix, $S^{[1]}$ is a $d \times d$ real diagonal matrix that holds the singular values of
$\Psi^{[1]}$ and $V^{[1]}$ is a $d^{L-1} \times d$ column orthogonal matrix, i.e.,
$V^{\dagger\,[1]}V^{[1]} = \I_d$.\\
Now one can define the elements of the
$d\times d^{L-1}$ matrix $S^{[1]} V^{\dagger\,[1]}$ as elements of a new
$d^2 \times d^{L-2}$ matrix $\Psi^{[2]}$
\begin{align}
  (S^{[1]}V^{\dagger\,[1]})_{\alpha_1,(\sigma_2,\ldots,\sigma_L)} :=
	\Psi^{[2]}_{(\alpha_1,\sigma_2),(\sigma_3,\ldots,\sigma_3)} \label{eq.def_Psi_i}
\end{align}
(with $\alpha_1 = 1, \ldots, d$), and apply the SVD again.
This process can be iterated for all quantum numbers $\sigma_i$.
In the end, one has
\begin{align}
  c_{\sigmas} &= \left( U^{[1]}_{\sigma_1} \cdots U^{[L-1]}_{\sigma_{L-1}} \cdot
	\Psi^{[L]}_{\sigma_L} \right) \ .
 \end{align}
As seen in Eq.\  \eqref{eq.def_Psi_i}, in each iteration, the quantum number $\sigma_i$ is shifted from the column index to the row index of $\Psi^{[i]}$.
Therefore, the matrices $U^{[i]}$ are of dimensions $d^i \times d^i$.
In order to carry out the matrix product, one has to select the right block (labeled by $\sigma_i$) from each $U^{[i]}$.
In other words, each $U^{[i]}$ and also the leftover $\Psi^{[L]}$ can be interpreted as a column vector of $d$ sub-matrices of dimension $d^{i-1} \times d^{i}$,
 which are indexed by $\sigma_i$
\begin{align}
  U^{[i]} = \left(\begin{matrix}
              A^{1} \\ \vdots \\ A^{d}
            \end{matrix}\right)
  \quad \text{with } A^{\sigma_i} \in \mathbb{C}^{d^{i-1} \times d^i}
	\label{eq.U_decomp} \ .
\end{align}
This quantity can also be seen as a tensor of order three
$\tens{A}^{\sigma_i}_{\alpha_{i-1},\alpha_{i}}$.
Eventually, a single coefficient $c_{\sigmas}$
is represented in the form
\begin{align}
  c_{\sigmas} &= \sum_{\alpha_1,\ldots,\alpha_L} \tens{A}^{\sigma_1}_{1,\alpha_1} \tens{A}^{\sigma_2}_{\alpha_1,\alpha_2} \cdots \tens{A}^{\sigma_{L-1}}_{\alpha_{L-1},\alpha_L} \tens{A}^{\sigma_L}_{\alpha_L,1}
	\nonumber \\
  &= A^{\sigma_1} \cdot A^{\sigma_2} \cdots A^{\sigma_L} \ .
	\label{eq.Csigma_matrix}
\end{align}
The index $\sigma_i$ denotes the physical state of the corresponding quantum number and is therefore referred to as physical index.
The index $1$ in $\tens{A}^{\sigma_1}$ and $\tens{A}^{\sigma_L}$ implies that these objects are vectors. Finally, the entire state reads
\begin{equation}
  \ket{\psi} = \sum_{\{\sigma_i\}} ( A^{\sigma_1} \cdots A^{\sigma_L} ) \,
	\ket{\{\sigma_i\}} \ .
	\label{eq.LC_MPS}
\end{equation}
Since each coefficient $c_{\sigmas}$ in Eq.\  \eqref{eq.Csigma_matrix} has the form of a product of $L$ matrices, the representation Eq.\  \eqref{eq.LC_MPS} is called a matrix product state.

By construction from the SVD, seen from the left end the matrices have increasing dimension: $1 \times d, d \times d^2, \ldots$ up to $i = \frac{L}{2}$ and are therefore distinct for different $\sigma_i$. Since we consider 1D chain models, this corresponds to open boundary conditions (OBC) where the position in the chain matters.
To implement periodic boundary conditions, all matrices have to be of the same dimension\footnote{In general, this dimension is $d^{L/2}\times d^{L/2}$. In special cases an exact MPS representation can be found even with $2\times 2$ matrices, e.g., for the 1D AKLT valence bond crystal \cite{affle87a,fanne92,schol11}.},
because the labeling of the sites can be shifted arbitrarily.
This is reflected in the cyclic property of the trace operation, which yields the proper scalar coefficient $c_{\sigmas}$ in this case  \cite{schol11}
\begin{align}
\label{eq.def_csigmas_tr}
  c_{\{\sigma_i\}} &= \Tr( A^{\sigma_1} \cdot A^{\sigma_2} \cdots A^{\sigma_L} ) \nonumber \\
   &= \Tr( A^{\sigma_2} \cdots A^{\sigma_L} \cdot A^{\sigma_1} ) \ .
\end{align}
Since Eq.\  \eqref{eq.Csigma_matrix} is a scalar expression, applying the trace does not change it and Eq.\  \eqref{eq.def_csigmas_tr} also holds for OBC.

In summary, generally the (maximum) dimension of the $A^{\sigma_i}$ grows as $d^{L/2}$ with $L$ and may vary with the lattice site depending on the boundary conditions.
However, for variational calculations, fixing all matrices to a given dimension $D \times D$  provides a way of truncating the Hilbert space which is systematic in the sense that it influences all bulk matrices in the same way. This $D$ is sometimes referred to as `bond dimension'.
Then it is more convenient to have the same dimension also at the ends of the system, regardless of the boundary conditions. To this end, the handling of the boundary conditions is shifted to two auxiliary systems of dimension $D$ located at both ends
of the chain with states $\ket{\alpha}$ and $\ket{\beta}$.
The corresponding matrices are vectors $\vec{\tilde a}^{\alpha\dagger}$ and
$\vec{\tilde b}^\beta$ of dimension $D$.
Putting everything together results in a very general ansatz for a MPS
\begin{subequations}
\label{eq.gen_MPS}
\begin{align}
  \ket{\psi} &= \sum_{\alpha,\beta} \sum_{\sigmas} c_{\sigmas}^{\alpha\beta} \ket{\sigmas} \ket{\alpha} \ket{\beta} \\
    &= \sum_{\alpha,\beta} \sum_{\sigmas} \Tr( \vec{\tilde a}^{\alpha\dagger} A^{\sigma_1} \cdots A^{\sigma_L} \vec{\tilde b}^\beta ) \, \ket{\sigmas} \ket{\alpha} \ket{\beta} \ .
	\label{eq.gen_MPS_tr}
\end{align}
\end{subequations}
Note, that in this representation the trace operation is redundant.
However, it is still helpful in understanding the way matrix elements and overlaps are computed in the thermodynamic limit, therefore, it is kept in the notation.
Another commonly used notation hides the boundary conditions in a boundary operator $Q$ in terms of which the MPS ansatz reads
\begin{align}
  \ket{\psi} &= \sum_{\sigmas} \Tr( Q A^{\sigma_1} \cdots A^{\sigma_L} ) \, \ket{\sigmas}
\end{align}
where the trace operation is then required to make the coefficients scalars.

Note that the MPS representation Eq.\  \eqref{eq.LC_MPS} is never unique.
\red{The construction starting the SVDs from the left side described in this section yields the so called \emph{left canonical} form of a MPS.
One could equally well start the decomposition from the right side or from both sides simultaneously meeting somewhere in the middle of the chain.
This would yield different matrices $A^{\sigma_i}$.
These canonical forms are very special representations because
there are many gauge degrees of freedom generally.}
Between any two matrix sets $A^{\sigma_i},\ A^{\sigma_{i+1}}$ one can always introduce an invertible matrix $X$ such that
\begin{align}
  c_{\sigmas} &= \Tr( A^{\sigma_1} \cdots A^{\sigma_i} \I A^{\sigma_{i+1}} \cdots A^{\sigma_L} ) \nonumber \\
  &= \Tr( A^{\sigma_1} \cdots (A^{\sigma_i} X_i) (X_i^{-1} A^{\sigma_{i+1}}) \cdots A^{\sigma_L} ) \nonumber \\
  &= \Tr( A^{\sigma_1} \cdots \tilde A^{\sigma_i} \tilde A^{\sigma_{i+1}} \cdots A^{\sigma_L} ) \label{eq.gauge}
\end{align}
which changes the adjacent matrices but leaves the coefficient $c_\sigmas$ unchanged.
Thus, equality of two states $\ket{\psi_1} = \ket{\psi_2}$ does not imply equality of their respective MPS matrix sets.
Therefore we understand the equality of two matrix sets $A^{\sigma_i}$ and $\tilde A^{\sigma_i}$ up to such a gauge transformation and as a shorthand meaning both sets represent the same state.

\subsection{Local operators}
\label{subsec:MPOs}

\begin{figure*}[t]
\normalsize
\begin{subequations}
\begin{align}
  \bracket{\psi_0}{\psi_0} &= \sum_{\alpha',\alpha,\beta',\beta} \sum_{\{s_i\}} \sum_{\{s_i'\}}
    \vec{\tilde a}^{\alpha' T} A^{s_1'\ast} \cdots A^{s_L'\ast} \vec{\tilde b}^{\beta' \ast} \,
    \vec{\tilde a}^{\alpha \dagger} A^{s_1} \cdots A^{s_L} \vec{\tilde b}^{\beta}
    \left\langle s_1',\ldots,s_L'|s_1,\ldots,s_L\right\rangle \bracket{\alpha'}{\alpha} \bracket{\beta'}{\beta} \tag{29a} \\
  &= \sum_{\alpha,\beta} \sum_{\{s_i\}} \Tr( \vec{\tilde a}^{\alpha T} A^{s_1'\,\ast} \cdots A^{s_L'\,\ast} \vec{\tilde b}^{\beta\ast} )
      \Tr( \vec{\tilde a}^{\alpha \dagger} Q A^{s_1} \cdots A^{s_L} \vec{\tilde b}^{\beta} )  \tag{29b} \\
  &= \sum_{\alpha,\beta} \sum_{\{s_i\}}
      \Tr\left[ (\vec{\tilde a}^{\alpha T} \otimes \vec{\tilde a}^{\alpha \dagger})
      ( A^{s_1\ast} \otimes A^{s_1} ) \cdots ( A^{s_L\ast} \otimes A^{s_L} )
      (\vec{\tilde b}^{\beta \ast} \otimes \vec{\tilde b}^{\beta}) \right]  \tag{29c} \\
  &= \Tr\Biggl[ \vec{a}^\dagger
    \underbrace{\left( \sum_{s_1 = 1}^{d} A^{s_1\ast} \otimes A^{s_1} \right)}_{=: T_1} \ \cdots\
    \underbrace{\left( \sum_{s_j = 1}^{d} A^{s_j\ast} \otimes A^{s_j} \right)}_{=: T_j} \ \cdots\
    \underbrace{\left( \sum_{s_L = 1}^{d} A^{s_L\ast} \otimes A^{s_L} \right)}_{=: T_L} \vec{b} \Biggr]  \tag{29d} \label{eq.def_T}
\end{align}
\end{subequations}
\addtocounter{equation}{-1}
\hrule
\end{figure*}

Having defined the matrix product representation of quantum mechanical states, a compatible definition of operators is introduced as well.
Analogous to a state being defined by its expansion coefficients with respect to the basis $\ket{\{\sigma_i\}}$, an operator $\hat O$  can be defined by its matrix elements
\begin{align}
  \bra{\{\sigma_i'\}} \hat O \ket{\{\sigma_i\}} &= \Tr( W^{\sigma_1'\sigma_1} \cdots W^{\sigma_L'\sigma_L} )
	\nonumber \\
  &= \sum_{\alpha_1,\ldots,\alpha_L} \tens{W}^{\sigma_1'\sigma_1}_{\alpha_L,\alpha_1} \cdots \tens{W}^{\sigma_L'\sigma_L}_{\alpha_{L-1},\alpha_L} \label{eq.def_Wsps}
\end{align}
where the quantities $\tens{W}^{\sigma_i'\sigma_i}_{\alpha_{i-1},\alpha_{i}}$ are tensors of order $4$ and for given $\sigma_i', \sigma_i$ the $W^{\sigma_i'\sigma_i}$ represents a matrix.
The general derivation and treatment of these objects is called matrix product operator (MPO) formalism  \cite{schol11}.

Generic Hamiltonians \eqref{eq.H_TFIM} consist of terms acting only on a small number of lattice sites. In the TFIM  one or two sites are involved.
This simplifies the general definition in Eq.\  \eqref{eq.def_Wsps}.
Let $\hat O$ be an operator that is the identity everywhere except at site $j$,
i.e., $\hat O$ is a single-site operator.
Then its matrix elements with respect to two MPS are given by
\begin{align}
 \bra{\phi} \hat O \ket{\psi}
      &= \sum_{\alpha\beta} \Tr\left[ \left( \vec{\tilde a}_\alpha^{\dagger \ast} \otimes \vec{\tilde a}_\alpha^{\dagger} \right)
	  \left( \sum_{\sigma_1} F^{\sigma_1 \ast} \otimes A^{\sigma_1} \right)
	  \right. \nonumber \\
      & \phantom{=} \left. \cdots \
	\left( \sum_{\sigma_j \sigma_j'} W^{\sigma_j'\sigma_j} F^{\sigma_j'\ast} \otimes A^{\sigma_j} \right)
      \cdots
      \left( \vec{\tilde b}_\beta^\ast \otimes \vec{\tilde b}_\beta \right) \right] \label{eq.apply_local_MPO}
\end{align}
\addtocounter{equation}{1}
where the matrices $F^{\sigma_i}$ describe the state $\ket{\phi}$,
the matrices $A^{\sigma_i}$  the state $\ket{\psi}$,
and $\otimes$ denotes the Kronecker product.
In the local Hilbert space of the single site $j$ the $W^{\sigma_j'\sigma_j}$ are just the elements of the matrix representation of $\hat O$, i.e., scalars.
This scheme readily extends to operators that are products of a finite number of
single-site operators (see Eq. \eqref{eq.gse_per_site}).

\subsection{Thermodynamic limit (iMPS)}
\label{subsec:TDL}

Let us consider the case, where the local Hilbert spaces at all sites
refer to locally identical spin degrees of freedom in a spin chain model such as the one defined in Eq.\  \eqref{eq.H_TFIM}.
The labels $\sigma_i$ run everywhere over the same set of values.
Furthermore, we assume that the Hamiltonian acting on these degrees of
freedom is the same at each site.
Then, the chain is translationally invariant in the thermodynamic limit $L \to \infty$.

\red{Given translational invariance, it is plausible to assume that a uniform MPS representation exists for separable ground states, i.e., states that can be separated in two blocks by a Schmidt decomposition \cite{fanne92}.
This means that all matrices $A^{\sigma_i}$ can be chosen the same.}
Such a state also results as a fixed point in the infinite system DMRG algorithm
\cite{ostlu95,romme97} and is called an iMPS.
The uniform ground state matrices will be labeled $A^s$ henceforth.
Next, we consider the norm of the ground state in Eq.\  (29).

\begin{center}
  \emph{see equation (29) above.}\\
\end{center}

In Eq.\  \eqref{eq.def_T} we defined the boundary vectors $\vec{a}^\dagger := \sum_{\alpha} \vec{\tilde a}^{\alpha T} \otimes \vec{\tilde a}^{\alpha\dagger}$ and
$\vec{b} := \sum_{\beta} \vec{\tilde b}^{\beta\ast} \otimes \vec{\tilde b}^{\beta}$.
The object $T$, which is also defined in Eq.\  \eqref{eq.def_T}, is called transfer operator or transfer matrix \cite{onsag44}.
Because the $A^{s_i}$ are the same at each site the transfer matrix is also uniform:
$T_i = T \ \forall i$.

The trace operation is redundant and used only to motivate the Kronecker product structure because of the identity $\Tr(A)\Tr(B) = \Tr(A\otimes B)$. Finally, the norm can be cast in the form
\begin{align}
\label{eq.norm_no_Q}
  \bracket{\psi}{\psi} &= \vec{a}^\dagger( T^{\dagger})^{\frac{L}{2}} T^{\frac{L}{2}} \vec{b} \ ,
\end{align}
which explains the name transfer matrix: If at some site $\vec{b}$ represents the right end of the chain the application of $T$ transfers the this chain end by one site to the left, i.e., it adds the next site.

From the definitions in Eq.\  \eqref{eq.def_T} it is obvious, that the transfer
matrix $T$ is of dimension $D^2 \times D^2$ and the vectors $\vec{a}$ and $\vec{b}$
are of dimension $D^2$. They can also be interpreted to be $D \times D$ matrices
$a$ and $b$ by filling such a matrix from top to bottom and left to right with the vector components
\begin{align}
\label{eq.vec2mat}
  \vec a = \left( \begin{matrix}
              \vec a_1 \\ \vdots \\ \vec a_D
             \end{matrix} \right)
  \quad \mapsto \quad
  a = \left( \begin{matrix}
        \vec a_1 & \ldots & \vec a_D
      \end{matrix} \right)
\end{align}
where the $\vec a_i$ are $D$ dimensional column vectors.
In this notation, the standard scalar product in $\mathbb{C}^{D^2}$ reads
\begin{equation}
  (a,b) = \Tr( a^\dagger b ) \ .
	\label{eq.matrix_sp}
\end{equation}

The application of $T$ to $b$ or of $T^\dagger$ to $a$ is also very concise
if $a$ and $b$ are denoted as matrices
\begin{subequations}
\label{eq.apply_T}
\begin{align}
  T[b] &= \sum_{s=1}^{d} A^{s} b \, A^{s \dagger} \\
  T^\dagger[a] &= \sum_{s=1}^{d} A^{s \dagger} a \, A^{s}
\end{align}
\end{subequations}
yielding again $D \times D$ matrices.
Note that, if $2d < D$, this evaluation of these expressions is computationally
more efficient than multiplying a $D^2 \times D^2$ matrix to a $D^2$ dimensional vector.

Let $\mu_i$ be the eigenvalues, $v_i$ the corresponding right eigenvectors (or eigenmatrices in the notation \eqref{eq.vec2mat})
of $T$, and $u_i$ the left eigenvectors, which are also the co-vectors of the $v_i$.
If $T$ is hermitian, $v_i = u_i$ holds. But this is generally not the case.
We consider the decomposition of $b$ into the $v_i$
\begin{align}
  b &= \sum_i (u_i, b) \, v_i =: \sum_i \beta_i v_i
	\nonumber \\
  \Rightarrow \quad T[b] &= \sum_i \beta_i \mu_i v_i \label{eq.T_ev_decomp} \ .
\end{align}
By the same argument that supports the power method for finding eigenvalues one
realizes that for very large $L$  \eqref{eq.norm_no_Q} implies
\begin{align}
  \frac{\bracket{\psi}{\psi}}{\mu_0^L} = \alpha_0^\ast u_0^\dagger \, v_0 \beta_0 \label{eq.norm_GS}
\end{align}
where $\mu_0$ is the largest eigenvalue in absolute value of $T$ and $u_0$ and $v_0$ are the corresponding left and right eigenvectors. This holds under the two conditions that the overlaps $\alpha_0 = \bracket{a}{v_0}$ and $\beta_0 = \bracket{b}{u_0}$ are finite and that $|\mu_0|$ is unique, i.e., there is no other eigenvalue
of the same absolute value.

\red{\emph{If} these two conditions are met, the explicit form of $\vec{a}$ and $\vec{b}$ and thus the boundary conditions they describe are irrelevant.
From the physical point of view, this is understood for
correlations of finite correlation length. Then, the behavior in the bulk of
the infinite system does not depend on the boundary conditions.}

\red{Note that the conclusion on the irrelevance of the
boundary conditions does not hold for degenerate ground states
where the boundary conditions may indeed influence the state in the bulk (see Appendix \ref{app:gs_degeneracy}).
Then the exact transfer matrix $T$ generically displays two or more eigenvalues of the same absolute value. For a physical example we refer to the Majumdar-Ghosh model \cite{majum69a,majum69b,majum69c}.
In its ground state, two spin-$\frac{1}{2}$ couple into a singlet state either on the odd or on the even bonds which leads to a two-fold ground state degeneracy
in the infinite system. This degeneracy is broken if there is a boundary:
If there is a boundary, the realized ground state favors a singlet on
the last bond in order to avoid a dangling spin.
In the corresponding analytical
transfer matrix we observe a two-fold degeneracy of $|\mu_0|$,
i.e., the above mentioned conditions are not met.}

\red{In the Ising phase, the TFIM also has a two-fold degenerate ground state.
As opposed to the Majumdar-Ghosh model, however, it does not have an exact iMPS representation at finite $D$. We observe that the
ground state search produces either one ground state or the other.
The superposition of both cannot be captured well by the MPS ansatz.
Around each of the two ground states, the above stated conditions hold and
$|\mu_0|$ is unique so that we may omit the boundary vectors $\vec{a}$ and
$\vec{b}$ from the notation unless stated otherwise.
In this sense, the description of the system  reduces to computing $\mu_0$, $v_0$ and $u_0$ and we will call $v_0$ and $u_0$ the boundary matrices.}

Once $\mu_0$ is known, $A^s$ can always be rescaled such that $|\mu_0| = 1$ and \eqref{eq.norm_GS} stays finite for $L \to \infty$.
If $\mu_0$ is positive, it can be rescaled to $\mu_0 \equiv 1$, otherwise, a phase factor remains\footnote{In this case, for every explicitly applied single-site operator (including $T$) the resulting matrix needs to be devided by $\mu_0$ to account for the phase factor. See Ref. \cite{ueda11} for a more detailed discussion of degenerate $\mu_0$.}.
This scaling for $\mu_0 \in \mathbb{R}$ is implied in the sequel.

Moreover, any scalar multiple of an eigenvector is also an eigenvector.
Therefore, $u_0$ and $v_0$ can be rescaled such that $(\alpha_0 u_0, \beta_0 v_0) = 1$.
These rescaled eigenvectors are labeled $u$ and $v$ and will be used from here on.
Due to the gauge freedom, see Eq.\  \eqref{eq.gauge}), one can find a gauge for $A^s$ such that either $v$ or $u$ equals the identity.
Then the other eigenmatrix is a diagonal matrix with non-negativ eigenvalues
and unit trace. It corresponds to the reduced density matrix appearing in DMRG.
This gauge has some advantages, see Appendix \ref{app:gs_search} for details), and is therefore the representation of choice.
For further details of this \emph{canonical} form of the infinite-size MPS (iMPS) see Ref. \cite{orus14}
\footnote{Ref. \cite{orus14} uses a composite representation $\{\Gamma, \Lambda\}$ of
the MPS where $\Gamma$ is a rank $3$ tensor that lives on the sites and $\Lambda$ is a
diagonal matrix that lives on the bonds. It holds the Schmidt coefficients of the
Schmidt decomposition across the bond. This representation can be obtained from the
$A^s$ tensors by computing the SVD $U = W \Lambda V^\dagger$ and setting
$\Gamma^s = V^\dagger W^s$. Here $U$ is the matrix $U^{[i]}$ from Eq.\
\eqref{eq.U_decomp} and the $W^s$ are the blocks of $W$ defined analogously to the
matrices $A^{\sigma_i}$. The canonical $A^s$ is obtained by $A^s = \tilde\Gamma^s \tilde\Lambda$ where $\{\tilde\Gamma,\tilde\Lambda\}$ is the canonical composite representation.}.

If not stated otherwise, we henceforth consider a non-degenerate tranfer matrix $T$
with unique $\mu_0=1$ after appropriate rescaling.
We consider  the single-site operator $\hat O$ from Eq.\
\eqref{eq.apply_local_MPO} to be the identity with the local matrix representation
$\I_d$.  Using Eq.\  \eqref{eq.norm_no_Q} we calculate the ground state expectation value in the thermodynamic limit
\begin{subequations}
\begin{align}
  \bra{\psi_0} \I \ket{\psi_0} &= \vec{a}^\dagger (T^\dagger)^{\frac{L}{2}} \left( \sum_{s,s'} (\I_d)_{ss'} A^{s'\ast} \otimes A^{s} \right) T^{\frac{L}{2}} \vec{b}
	\\
  &= \vec{u}^\dagger \left( \sum_{s} A^{s\ast} \otimes A^{s} \right) \vec{v}
	\\
  &= \vec{u}^\dagger T \vec{v} = (u,T[v]) = 1 \ .
\end{align}
\end{subequations}
In this sense, $T$ can also be perceived as an identity operation at one site
\begin{align}
  T[v] &= \sum_{s} A^{s} v \, A^{s\dagger} =  \sum_{s, s'} \delta_{ss'} A^{s'} v \, A^{s \dagger}
	\nonumber \\
  &= \sum_{s,s'} (\I_d)_{ss'} A^{s'} v \, A^{s \dagger} =: \I^{(A,A)}[v] = v \ .
\end{align}

The scheme in Eq.\  \eqref{eq.apply_T} extends to nontrivial operators
straightforwardly as follows
\begin{subequations}
\label{eq.apply_op}
\begin{align}
  \hat O^{(B,A)}[v] &= \sum_{s,s'} O_{ss'} \, A^{s'} v \, B^{s \dagger}
	\\
  \hat O^{\dagger (B,A)}[u] &= \sum_{ss'} O^\dagger_{ss'} \, A^{s\dagger} u \, B^{s'} \ .
\end{align}
\end{subequations}
As an example for the application of a local MPOs let us consider a single term
\begin{align}
  h_i = - \Gamma \Sz_i - J \Sx_i \Sx_{i+1} \label{eq.def_hi}
\end{align}
of the Hamiltonian \eqref{eq.H_TFIM}.
Applying the scheme in Eq.\  \eqref{eq.apply_local_MPO} using \eqref{eq.apply_op} yields
\begin{subequations}
\label{eq.gse_per_site}
\begin{align}
  \bra{\phi} h_i \ket{\psi} =
    & -\Gamma \bra{\phi} \Sz_i \ket{\psi} - J \bra{\phi} \Sx_i \Sx_{i+1} \ket{\psi} \\
  = \, & - \Gamma (\bar u, S^{z\,(F,A)}[\bar v]) \ + \label{eq.gse_TF} \\
    \, & - J (\bar u, S^{x\,(F,A)}[  S^{x\,(F,A)}[\bar v] ]) \label{eq.gse_Ising}
\end{align}
\end{subequations}
where $\bar v$ and $\bar u$ are the eigenvectors of $\bar T = \sum_s F^{s \ast} \otimes A^s$.

This concludes the brief formal and technical review of the matrix product
formalism. Below we turn to its application to the TFIM and to the
construction of local creation and annihilation operators.

\section{Ground state energy and dispersion}
\label{sec:GSE}

In this section, we describe one of several ways to obtain a uniform iMPS
representation of the ground state
and how to calculate the dispersion of a single quasi-particle in the system.

\subsection{Ground state search}

Starting from the ansatz \eqref{eq.gen_MPS}, finding the ground state energy is a variational problem in the coefficients of the ground state matrices $A^s$
\begin{align}
  \frac{E_0}{L} \leq \min_{\{A^s\}} \frac{\bra{\psi_0(A^s)} h_i \ket{\psi_0(A^s)}}{\bracket{\psi_0(A^s)}{\psi_0(A^s)}} \label{eq.Rayleigh_Ritz}
\end{align}
where $h_i$ is the local term of the Hamiltonian defined in Eq.\  \eqref{eq.def_hi}.
There are various methods of finding an optimal $A^s$ for given $D$\red{.}

Eq.\  \eqref{eq.Rayleigh_Ritz} is a highly nonlinear function in the elements of
$A^s$. Thus, for its minimization, one may think to resort to any multi-dimensional minimizer that does not rely on derivatives.
But the convergence of them is usually slow.
\red{Another non-variational possibility for the ground
state search is the imaginary time evolution in Vidal's iTEBD \cite{vidal07}
which, however, is also found to much slower than MPS-based iDMRG \cite{mccull08}}.

Alternatively, one may use an iterative approach.
In each step, only the elements of the matrcies $B^s$ at a single site are varied, all other sites are kept at a fixed $A^s$.
Then, the matrix $B^s$ with the lowest ``local energy'' $\epsilon_0$ is adopted everywhere as improved guess for $A^s$ and this process is iterated untill convergence $B^s_0 = A^s$ is reached within some numerical tolerance.

Let $\ket{\psi(A^s,B^s)}$ be the state that has $A^s$ matrices everywhere except at site $i=0$ where the $B^s$ matrices are inserted instead.
This insertion breaks the uniformity of the state.
Therefore, its energy is no longer given by a multiple of the expectation value of $h_i$.
Instead the full Hamiltonian $\mathcal{H} = \sum_i h_i$ has to be taken into account.
The minimization problem in terms of the elements of $B^s$  reads
\begin{align}
  \epsilon = \frac{\bra{\psi(A^s,B^s)} \left( \sum_i h_i - E(A^s) \right) \ket{\psi(A^s,B^s)}}{\bracket{\psi(A^s,B^s)}{\psi(A^s,B^s)}}
	\label{eq.min_Bs}
\end{align}
where $E(A^s)$ is the energy per site of the uniform state that has only $A^s$ matrices.
This is also the best estimate for the ground state energy per site $E_0/L$ at each step
of the iteration. We subtract it in order to avoid extensive contributions.

Both the numerator and the denominator of Eq.\  \eqref{eq.min_Bs} are bilinear forms in the $d\cdot D^2$-dimensional vector $\vec B$ that holds all the elements of the matrix set $B^s$. Looking for minima in Eq.\  \eqref{eq.min_Bs} means looking for roots in its derivative with respect to $\vec B^\dagger$. It is well-established  that for bilinear forms the roots of this derivative  amount up to the generalized eigenvalue problem (EVP)
\begin{subequations}
\label{eq:evp}
\begin{align}
  \eqref{eq.min_Bs} \quad &\Leftrightarrow \quad \vec B^\dagger M(\mathcal{H},A^s) \vec B = \epsilon \vec B^\dagger N(A^s) \vec B
	\\
  &\rightarrow \quad M(\mathcal{H},A^s) \vec B = \epsilon N(A^s) \vec B \label{eq.GSE_EVP} \ .
\end{align}
\end{subequations}
Note that the matrices $M$ and $N$ are both Hermitian by construction.
For details on the ground state search algorithm, we refer the reader to
appendix \ref{app:gs_search}.

There is no rigorous proof that adopting the local minimum $B^s_0$, that is found
from the minimization at one site, at all sites will  lower the total  energy.
But empirically it is found to be the case if the initial guess for the
$A^s$ is not too far away from an optimal $A^s$. Moreover, in practice we
adopt a line-search algorithm between $B^s_0$ and the former $A^s$ to stabilize
the minimization, see appendix \ref{ss:fine-tuning}.

\begin{figure}
\centering
\includegraphics[width=\columnwidth]{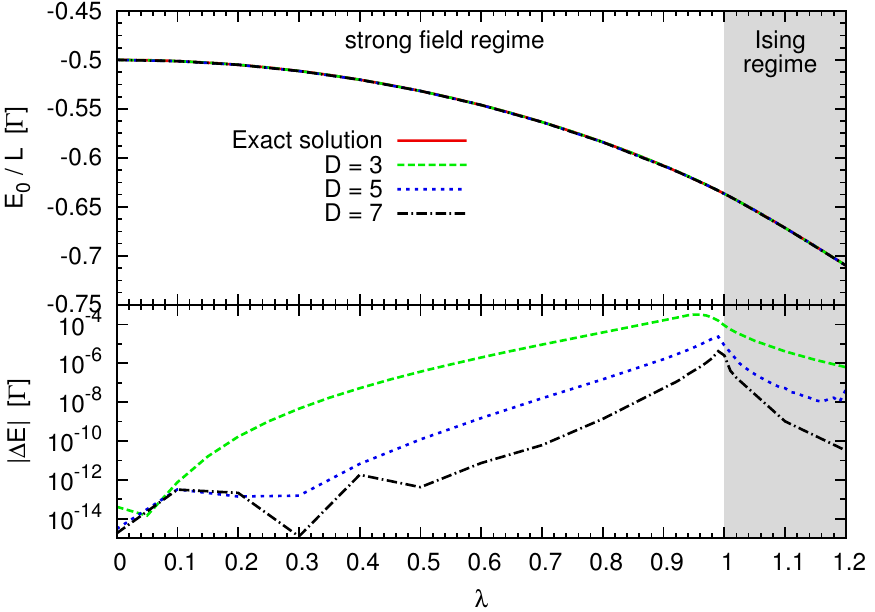}
\caption{(Color online) Upper panel: Ground state energy per site $E_0/L$ as
function of $\lambda$.
Comparison of the exact result \eqref{eq.E0_TFIM}  to results from iMPS
calculations with various bond dimensions $D$. The lower panel shows the deviation
$|\Delta E| = |E_0/L - E_\text{0,exact}|$. The critical point is located at
$\lambda = 1$ and the shaded area to its right marks the Ising regime with
two-fold degenerate ground state.}
\label{plot.GSE}
\end{figure}

Results for the ground state energy of the TFIM are depicted in Fig.\ \ref{plot.GSE}.
The agreement is extremely good in view the low matrix dimension.
There is a clear maximum in the deviation from the exact result, close to the location of the phase transition.
The parameter value of the largest deviation is found to be
below the true critical values $\lambda < \lambda_c=1$, but quickly approaches it as $D$ grows.

As explained for instance in Ref.\ \cite{orus14}, the MPS formalism inherently implies exponentially
decaying correlations because of the finite, bounded amount of entanglement which can be represented.
The length scale of the decay of these correlations, the correlation length, is determined by the second largest magnitude eigenvalue $\mu_1$ of $T$.
To see this, consider the application of $T$ to a matrix $b$ expanded in eigenmatrices
$v_i$ of $T$ in Eq.\  \eqref{eq.T_ev_decomp}.
Assuming a non-degenerate spectrum of $T$ and $\mu_0 = 1$ the subleading term is
$\beta_1 \mu_1 v_1$ with $|\mu_1| < 1$.
This term determines the rate at which $T^j[b]$ converges to $\beta_0 v_0$
\begin{align}
  T^j[b] \approx \beta_0 v_0 + \mu_1^j \beta_1 v_1 \ .
\end{align}
Therefore, the correlation length $\xi_T$ captured by $T$ is given as
\begin{align}
  \xi_T = - \frac{1}{|\ln\mu_1|}
	\label{eq.xi_T} \ .
\end{align}

\begin{figure}
\centering
\includegraphics[width=\columnwidth]{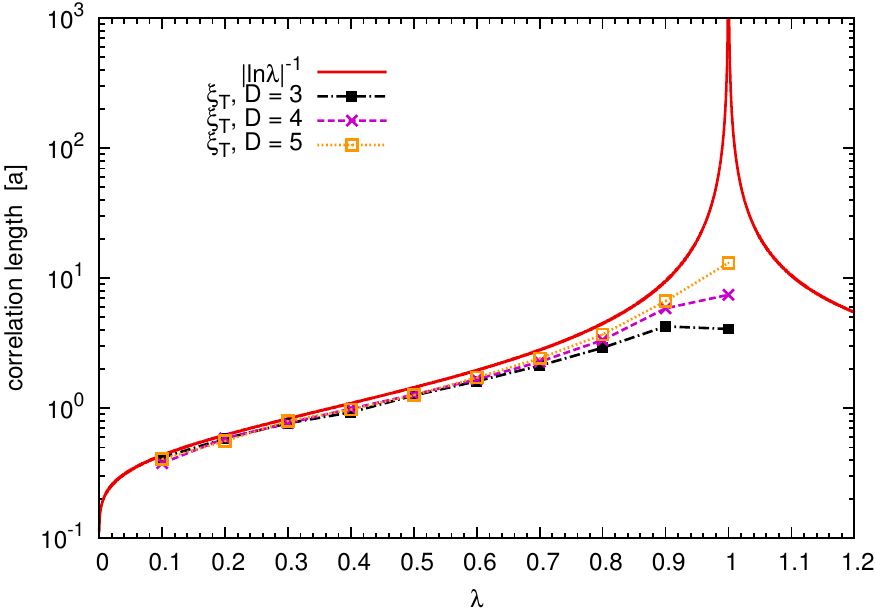}
\caption{(Color online) The correlation length $\xi_T$ as computed from the second largest EV of the transfer matrix, see Eq.\  \eqref{eq.xi_T}, compared to the exact expression $\xi = |\ln\lambda|^{-1}$ from \eqref{eq.xi_TFIM}.}
\label{plot.ZetaXi}
\end{figure}

Figure \ref{plot.ZetaXi} displays $\xi_T$ for various matrix dimensions $D$ in comparison to the exact expression Eq.\  \eqref{eq.xi_TFIM}.
Especially close to criticality a larger matrix dimension is required to improve the agreement.
\red{Since this is a proof-of-concept study, the computations were carried out on laptop computers and workstations. Therefore, we restricted the bond dimension $D$ to low values to keep the runtime short. Bond dimensions of several hundred are possible, but then the calculations take considerably more time.}

\subsection{Dispersion}

For simplicity, we focus here on the regime where the ground state is unique.
In Appendix \ref{app:gs_degeneracy} the changes for degenerate ground
state are summarized.
\red{The approach is also described in Ref.\ \cite{Haegeman2013b}.}

If the ground state is unique, the eigenvectors $\vec B_{\alpha > 0}$ of
\eqref{eq.GSE_EVP} with higher local energy $\epsilon_{\alpha>0}$ describe excitations of the system. Let
\begin{align}
  \ket{\psi^\alpha_j} = \sum_{\{s_i\}} \Tr(A^{s_1} \cdots A^{s_{j-1}} B^{s_j}_\alpha A^{s_{j+1}} \cdots A^{s_L}) \ket{\{s_i\}}
	\label{eq.def_Psi_ARS}
\end{align}
be the state that has ground state matrices everywhere except at site $j$ where a
$B^s_\alpha$ matrix is inserted instead.
By construction, these states are orthogonal to the ground state
\begin{align}
  \bracket{\psi_0}{\psi^\alpha_j} = (u,\I^{(A,B_\alpha)}[v]) = \vec A^{\dagger} N \vec B_\alpha = 0 \ .
\end{align}
The same holds for states with different $\alpha$ if the $B^s_\alpha$ are at the same site since
\begin{align}
  \bracket{\psi^\alpha_j}{\psi^\beta_j} = \vec B^{\dagger}_\alpha N \vec B_\beta \propto \delta_{\alpha\beta}
\end{align}
because they result from the same generalized EVP \eqref{eq:evp}
with Hermitian matrices $M$ and $N$.

But if $B^s_\alpha$ and $B^s_\beta$ are inserted at different sites $j' \neq j$ the corresponding states will not be orthogonal.
If we want to view the insertion $A^s \to B^s_\alpha$ as the effect of a creation operator
we need that the excitations at different sites are mutually orthogonal.
How can we solve this issue? We achieve orthogonality by resorting to the
the construction of Wannier states known from solid state text books.
One takes a detour via the Fourier transform since the resulting momentum eigenstates
\begin{align}
  \ket{\psi^\alpha_q} := \frac{1}{\sqrt{L}} \sum_j e^{-iqj} \ket{\psi^\alpha_j} \label{eq.def_Psi_AQS}
\end{align}
are known to be orthogonal in momentum space $\bracket{\psi_q^\alpha}{\psi_{q'}^\beta} \propto \delta_{qq'}$ because they refer to different eigenvalues under discrete translations.
The restriction of all momenta to the first Brillouin zone is implied.

Exploiting translational invariance, the overlap of two momentum eigenstates can be computed as
\begin{subequations}
\label{eq.def_Nq}
\begin{align}
  \bracket{\psi_q^\alpha}{\psi_q^\beta} &= \frac{1}{L} \sum_{j,j'} e^{iqj'} e^{-iqj} \bracket{\psi^\alpha_{j'}}{\psi^\beta_j}
	\\
  &= \frac{1}{L} \sum_{j,j'} e^{iq(j'-j)} \bracket{\psi^\alpha_{j'-j}}{\psi^\beta_0}
	\\
  &= \sum_j e^{iqj} \bracket{\psi^\alpha_j}{\psi^\beta_0}
	\\
  &=: N_q^{\alpha\beta} \label{eq.def_Nq_ab}
\end{align}
\end{subequations}
where Eq.\  \eqref{eq.def_Nq} defines the matrix $N_q$ which is the metric tensor of the states $\ket{\psi_q^\alpha}$.
As seen in Eqs.\ \eqref{eq.def_Nq}, the normalization factor $\frac{1}{L}$ always cancels out in the computation of an expectation value, norm or overlap.
Thus the limit $L \to \infty$ does not pose any numerical problems.
As seen in the sequence of equalities \eqref{eq.def_Nq}, translational invariance allows us to assume that the ket-side matrices $B^s$ are always placed at site $0$.

The infinite sums over real space indices in the above equations may be seen
as insurmountable problem. But this is not the case because the
infinite sums converge exponentially.
To elucidate this point we look at the following limits.
Any $D \times D$ matrix $m$ (assuming $\mu_0 = 1$) can be decomposed according to Eq.\  \eqref{eq.T_ev_decomp}.
Applying the transfer matrix $T$ $j$-times yields
\begin{subequations}
\label{eq.Tv_convergence}
  \begin{align}
    T^j[m] &= \sum_i (u_i,m) T^j v_i
		\nonumber \\
    &= \sum_i (u_i,m) (\mu_i)^j v_i
		\\
    & \quad\quad \text{with } |\mu_i| < 1 \text{ for } i > 0
		\nonumber \\
    &\Rightarrow \quad \lim_{j\to\infty} T^j [m] = (u,m) \cdot v
		\\
    &\phantom{\Rightarrow} \quad \lim_{j\to\infty} T^{\dagger\, j} [m] = (m,v)
		\cdot u
  \end{align}
\end{subequations}
where the convergence to these limits is exponential in $j$
governed by the second largest absolute value $|\mu_1|$ of the eigenvalues of
$T$.

The overlap $\bracket{\psi^\alpha_j}{\psi^\beta_0}$ goes to zero for large $j$
\begin{subequations}
\label{eq.Nj_decay}
\begin{align}
  \bracket{\psi^\alpha_j}{\psi^\beta_0} &= (\I^{(B_\alpha,A)}[u], T^j[ \I^{(A,B_\beta)}[v]] ) \\
  &\approx ( \I^{(B_\alpha,A)}[u], (u, \I^{(A,B_\beta)}[v]) \cdot v )
	\label{eq:approx}
	\\
  & = \bracket{\psi_0}{\psi^\beta_0} \bracket{\psi^\alpha_j}{\psi_0}
	\\
  & = 0
\end{align}
\end{subequations}
where the approximation in \eqref{eq:approx} refers to the exponential convergence
established in Eqs.\ \eqref{eq.Tv_convergence}. In the last line we used
the local orthogonality $\bracket{\psi_0}{\psi^\alpha_0} = 0$.
The exponential convergence to zero justifies to trunate the Fourier series after a finite number $j_\text{max}$ of terms.

If the ground state search is converged, $\vec B_{\alpha=0}$ is the set of ground state matrices, i.e., $\ket{\psi_i^{\alpha=0}} \equiv \ket{\psi_0} \ \forall\ i$.
If the ground state matrices are established with sufficient numerical accuracy,
 all $\ket{\psi^{\alpha > 0}_q}$ are orthogonal to $\ket{\psi_0}$.
Thus, the dispersion relation can be found by a second variational calculation in the orthogonal complement of the ground state, i.e., in the sub space spanned by the
$\ket{\psi^{\alpha>0}_q}$
\begin{subequations}
\label{eq.dispersion_min}
\begin{align}
  \omega_q &\leq \min_{\vec v_q} \frac{\bra{\phi_q} (\mathcal{H} - E_0) \ket{\phi_q}}{\bracket{\phi_q}{\phi_q}}
	\\
  \ket{\phi_q} &:= \sum_{\alpha = 1}^{d\cdot D^2-1} v_q^\alpha \ket{\psi^\alpha_q} \label{eq.RR_dispersion} \ .
\end{align}
\end{subequations}
This leads to another generalized EVP
\begin{align}
\label{eq.dispersion_EVP}
  H_q \vec v_q &= \omega_q N_q \vec v_q
\end{align}
where $N_q$ is the matrix defined in Eq.\  \eqref{eq.def_Nq_ab} and $H_q$ is defined analogously
\begin{subequations}
\begin{align}
  H_q^{\alpha\beta} &:= \bra{\psi^\alpha_q} (\mathcal{H}-E_0) \ket{\psi^\beta_q}
	\label{eq.Hq_elements}
	\\
  &= \sum_{j,i} e^{iqj} h^{\alpha\beta}_{j,i}
	\\
  &:= \sum_{j} e^{iqj} \left[ \sum_i \bra{\psi^\alpha_j} \left(h_i - \frac{E_0}{L} \right) \ket{\psi^\beta_0} \right] \ .
\end{align}
\end{subequations}
The lowest eigenvalue $\omega^0_q$ is the best estimate for the one-particle dispersion at given $q$.

Haegeman et al.\ \cite{haege12} observed that there is always a number of choices
 $B^s_\alpha$ such that $\ket{\psi^\alpha_q} \equiv 0 \ \forall\ q$
 due to the gauge degrees of freedom stated in Eq.\  \eqref{eq.gauge} combined with translational invariance. Because of the associativity of the matrix product for any
$X \in \mathbb{C}^{n \times n}$ we have
\begin{align}
  \ket{\psi_{j-1}^R} &= \sum_{\{s_i\}} \Tr( \cdots (A^{s_{j-1}} X) \tilde A^{s_{j}} \cdots ) \ket{\{s_i\}} =
	\nonumber \\
  \ket{\psi_{j}^L} &= \sum_{\{s_i\}} \Tr( \cdots A^{s_{j-1}} (X \tilde A^{s_{j}}) \cdots ) \ket{\{s_i\}} \ .
\end{align}
Note that we allow here for the more general case where the ground state matrices
are different to the left and to the right of the inserted gauge matrix $X$.
This includes the  possibility of excitations of domain wall character where
one switches between degenerate ground states.

Let us define $B^s := e^{iq} A^s X - X \tilde A^s$ implying
\begin{subequations}
\label{eq.Hj_decay}
\begin{align}
  \ket{\psi_q} &= \frac{1}{\sqrt{L}} \sum_j e^{-iqj} \ket{\psi_j}
	\\
  &= \frac{1}{\sqrt{L}} \sum_j e^{-iqj} \left( e^{iq} \ket{\psi_j^R} - \ket{\psi_j^L} \right)
	\\
  &= \frac{1}{\sqrt{L}} \sum_j e^{-iqj} \left( \ket{\psi_{j-1}^R} - \ket{\psi_{j}^L} \right)
	\\
  &= 0
\end{align}
\end{subequations}
because a phase factor of $e^{iq}$ translates to a shift of the states in real space by one site to the left under the Fourier transformation.
The matrix $X$ has $D^2$ parameters.
Thus, for $q \neq 0$ or for $q = 0$ and $\tilde A^s \neq A^s$, the dimension of the space spanned by $\ket{\psi_q^\alpha}$ is reduced by $D^2$.
In other words, there are $D^2$ ``zero modes''.

For $q = 0$ and $\tilde A^s = A^s$, the choice of $X = \I$ results in $B^s = 0$ which makes $\ket{\psi_j}$ the null vector.
Thus the number of linearly independent zero modes is reduced to $D^2-1$ in this case.
Therefore, the metric tensor $N_q$ has a $D^2$ or $(D^2-1)$ dimensional null space.
In order to take this into account $H_q$ needs to be projected onto the non-zero eigenspace for computing $\omega_q$.
Within this non-zero eigenspace, the dispersion is found by solving the standard EVP
\begin{align}
  \sqrt{D_N}^{\,-1} V'^\dagger H_q V' \sqrt{D_N}^{\,-1} \, \vec v\,'_q = \omega_q \vec v\,'_q
	\label{eq.gen_to_std_EVP}
\end{align}
where the diagonal matrix $D_N$ holds the non-zero eigenvalues of $N_q$ and $V'$ the corresponding eigenvectors. The original $\vec v_q$ from Eq.\
\eqref{eq.dispersion_EVP} can be obtained as $\vec v_q = V' \vec v\,'_q$.

The computation of the matrix elements in Eq.\  \eqref{eq.Hq_elements} is the most time
consuming part of the calculation because the complete Hamiltonian acts on all lattice
sites and the Fourier coefficients have to be computed for many values of $j$.
But by the same argument as in Eq.\  \eqref{eq.Nj_decay}, the contributions
converge exponentially to zero if $|j| \gg 1$.
Let for instance $j \ll 0$, $i \gtrsim 0$. Then
\begin{subequations}
\begin{align}
  h^{\alpha\beta}_{j,i} &= \bra{\psi^\alpha_j} h_i \ket{\psi^\beta_0} - \frac{E_0}{L}\bracket{\psi^\alpha_j}{\psi^\beta_0}
	\\ \nonumber
  &= (\I^{(B_\alpha,A)}[ T^{\dagger\,|j-1|}[ \I^{(A,B_\beta)}[u] ] ],
	T^{i-1}[ h_i[v] ] )
	\\
  &\qquad\qquad - \frac{E_0}{L}\bracket{\psi^\alpha_j}{\psi^\beta_0}
	\\
  &\approx \left( \I^{(B_\alpha,A)}[(\I^{(A,B_\beta)}[u],v) \cdot u], T^{i-1}[ h_i[v] ] \right) \\
  &= \bracket{\psi_0}{\psi^\beta_0} (\I^{(B_\alpha,A)}[u],T^{i-1}[ h_i[v] ]) = 0 \ .
\end{align}
\end{subequations}
where the vanishing of the last expression holds in the limit $j\to \infty$.

For $|i| \gg 1$ we obtain similarly
\begin{align}
  h^{\alpha\beta}_{j,i} &= (\I^{(B_\alpha,A)}[T^{\dagger |j-1|}[\I^{(A,B_\beta)}[u]]],
	T^{i-1}[h_i[v]])
	\nonumber \\
  &\qquad\qquad - \frac{E_0}{L} \bracket{\psi^\alpha_j}{\psi^\beta_0}
	\nonumber \\
  &\approx \left(\I^{(B_\alpha,A)}[T^{\dagger |j-1|}[\I^{(A,B_\beta)}[u]]], v \cdot \frac{E_0}{L}\right)
	\nonumber \\
  &\qquad\qquad - \frac{E_0}{L} \bracket{\psi^\alpha_j}{\psi^\beta_0}
	\nonumber \\
  &= \bracket{\psi^\alpha_j}{\psi^\beta_0} \frac{E_0}{L} - \frac{E_0}{L} \bracket{\psi^\alpha_j}{\psi^\beta_0} = 0 \quad \text{for} \quad j\to\infty
\end{align}
where  $j < 0$ is assumed for simplicity.

\begin{figure}
\centering
\includegraphics[width=\columnwidth]{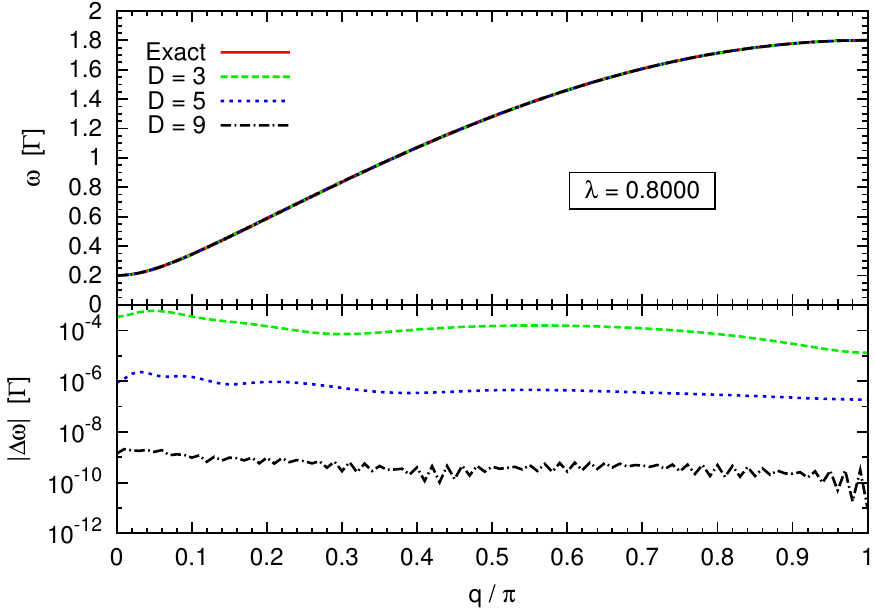}
\caption{(Color online) Dispersion for $\lambda = 0.8$ in the strong-field regime.
Comparison of the exact result \eqref{eq.Dispersion_TFIM} to results from iMPS calculations with various matrix dimensions $D$.}
\label{plot.Dispersion_sf}
\end{figure}

\begin{figure}
\centering
\includegraphics[width=\columnwidth]{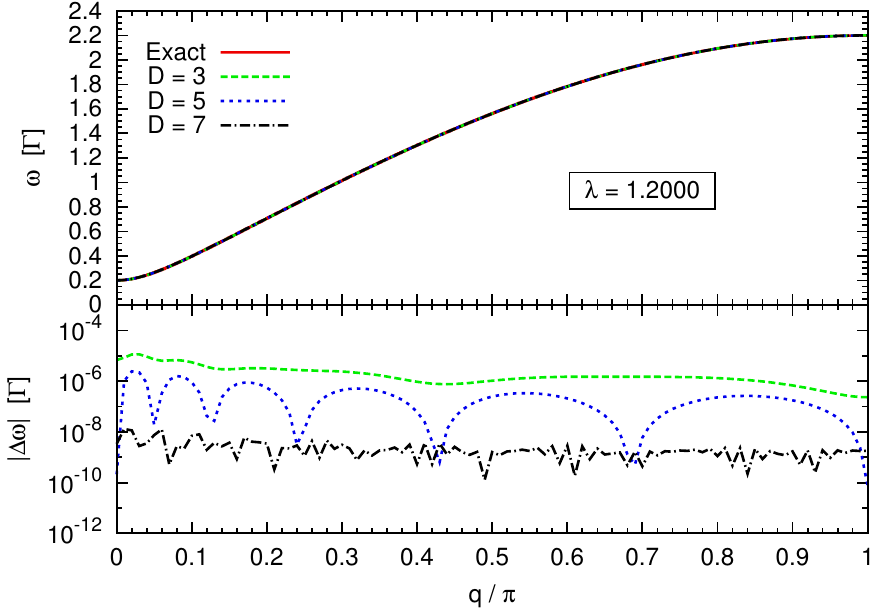}
\caption{(Color online) Dispersion  for $\lambda = 1.2$ in the Ising regime.
Comparison of the exact result \eqref{eq.Dispersion_TFIM} to results from iMPS calculations with various matrix dimensions $D$.}
\label{plot.Dispersion_is}
\end{figure}

\begin{figure}
\centering
\includegraphics[width=\columnwidth]{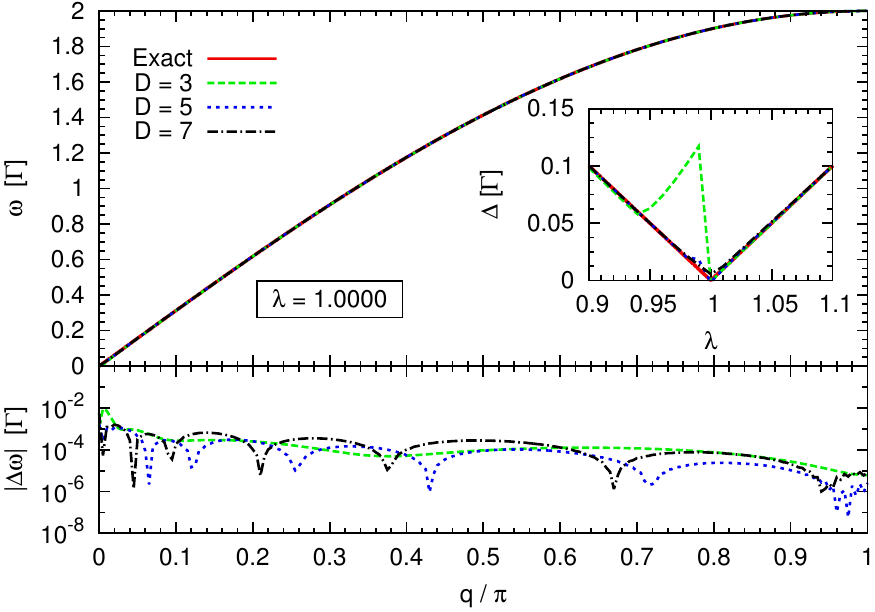}
\caption{(Color online) Dispersion  at $\lambda = 1$, i.e., at the quantum critical point. Comparison of the exact result \eqref{eq.Dispersion_TFIM} to results from iMPS calculations with various matrix dimensions $D$. The inset focuses on the gap as function of $\lambda$ around the critical point.}
\label{plot.Dispersion_crit}
\end{figure}

Figures \ref{plot.Dispersion_sf} through \ref{plot.Dispersion_crit} show the dispersion $\omega_q$ for various parameter values.
At $\lambda = 0.8$ and $\lambda = 1.2$, see Figs.\ \ref{plot.Dispersion_sf} and \ref{plot.Dispersion_is}, the agreement is very good, both in the strong field and in the Ising regimes, because the system is placed not too close to criticality.
The nice agreement illustrates that ground state degeneracy is handled very well by the advocated method.

Directly at the quantum critical point, see Fig.\ \ref{plot.Dispersion_crit}, the closing of the gap is difficult to capture numerically.
The reason is the diverging correlation length $\xi$, see \eqref{eq.xi_TFIM}.
The amount of entanglement that can be described by an MPS is bounded by the matrix dimension $D$.
Thus, no finite-dimensional MPS can completely describe a state with diverging correlation length.

The inset in Fig.\ \ref{plot.Dispersion_crit} depicts the gap $\Delta$ as function of
$\lambda$ in the vicinity of $\lambda = 1$. Note that the gap values are as low
as $10^{-6}\Gamma$ to $10^{-5}\Gamma$ in spite of the limited bond dimension.
The occurrence of a rather sharp minimum indicates a possible phase transition.
Note that this criterion is independent of a comparison to the exact result
and allows one to estimate the corresponding critical parameter value as well.

\section{Effective model}
\label{sec:eff_model}

As mentioned in Sect. \ref{sec:model}, in the strong-field limit $\lambda \to 0$ the elementary excitations are flips of single spins from the polarized ground state.
If the Ising interaction is switched on by a finite $\lambda$, such a flipped spin
acquires a virtual dressing which corresponds to a polarization cloud around
it. This means that the elementary excitations are no longer strictly local, but
``smeared out'' over a certain region on the chain. This concept has
been the basis of the CUTs in real space representation
\cite{knett00a,knett03a,yang11a,krull12}. The spatial extension of the
polarization cloud, i.e., of the smeared out region, is governed
by the correlation length. This can also be seen from
Eq.\ (4) in Ref.\ \cite{haege13a}.

In order to make progress in deriving effective models \eqref{eq.H_eff}
in terms of the elementary excitations we want to establish the key
ingredients of second quantization, namely the creation and annihilation
operator of an elementary excitation. Thus, it is
our objective in this section to explicitly derive a local creation
operator acting on the ground state. If we know the ground state (or a very
good numerical representation thereof) and we are able to
characterize the local excited states we can follow the route advocated
previously \cite{knett03a} to determine the effective model on the
bilinear level.

More work will be required for the determination of
decay terms \eqref{eq.H_decay} and two-particle interactions  \eqref{eq.H_int}.
To determine them, states with two quasi-particles at sites $i$ and $j$ must be properly defined.
This requires that they are normalized and two such states
$(i,j)$ and $(i',j')$ are orthogonal if $i\neq i'$ or $j\neq j'$.
Moreover, such two-particle states must fit to the one-particle states in the
sense that they decompose into the one-particle states for $|i-j|\to\infty$.
These issues set the roadmap of research, but they are beyond the scope
of the present article.

In order to construct a local creation operator, we consider the eigenvector
$\vec v_q$ of Eq.\  \eqref{eq.dispersion_EVP} that belongs to the
lowest eigenvalue which defines  the dispersion $\omega_q$.
Its components $v^\alpha_q$ describe how the states $\ket{\psi^\alpha_q}$ are
linearly combined to form an elementary excited state that satisfies
\begin{subequations}
\begin{align}
  &\bra{\phi_q} (\mathcal{H} - E_0) \ket{\phi_q} = \omega_q \\
  &\qquad \ket{\phi_q} = \sum_\alpha v^\alpha_q \ket{\psi^\alpha_q} =
	a^\dagger_q \ket{\psi_0}
	\label{eq.def_AdQ} \ .
\end{align}
\end{subequations}
This means that  $\ket{\phi_q}$ can be interpreted as a state, in which one quasi-particle of momentum $q$ has been created.
Taking the inverse Fourier transform of Eq.\  \eqref{eq.def_AdQ} one obtains an expression for the action of a local creation operator $a^\dagger_i$ on the ground state
\begin{subequations} \label{eq.def_Ad_rs}
\begin{align}
  &a_i^\dagger \ket{\psi_0} = \sum_{j,\alpha} v_j^\alpha \ket{\psi^\alpha_{i+j}}
      \label{eq.def_Ad_i} \\
  &\quad \text{with} \quad v_j^\alpha := \frac{1}{L} \sum_q v_q^\alpha e^{iqj}
      \label{eq.def_vj} \ .
\end{align}
\end{subequations}
This equation is the key element in advancing towards effective models
via MPS representations.

In the thermodynamic limit $q$ is a continuous variable and the sum in Eq.\  \eqref{eq.def_vj} becomes the integral over the Brillouin zone
\begin{align}
  v_j^\alpha := \frac{1}{2\pi} \int_{-\pi}^{\pi} v_q^\alpha \, e^{iqj} \, \d q
	\label{eq.def_Ad_i_int} \ .
\end{align}
Although numerical integration always comes down to summation at some point, the
continuous representation is advantageous for adaptive algorithms
because Eq.\  \eqref{eq.dispersion_EVP} can be evaluated at arbitrary values of $q$.
See Appendix \ref{app:creation_op} for comments and technical details on handling
$\vec{v}_q$.

Taking the sum over $\alpha$ first in Eq.\  \eqref{eq.def_Ad_i}
simplifies the numerical computation. Hence, we define a single matrix set
\begin{align}
  C^s_j := \sum_\alpha v_j^\alpha B^s_\alpha
	\label{eq.def_Csj}
\end{align}
to be inserted at distance $j$ from the center site $i$ of the particle created by
$a_i^\dagger$.
In this description the particle is represented by a number of matrices $\{C^s_j\}$ as follows
\begin{align}
  a^\dagger_i \ket{\psi_0} = \sum_{j=-j_\text{max}}^{j_\text{max}}
	\ket{\psi(C^s_j)_{i+j}}
	\label{eq.def_Ad_i_compact}
\end{align}
where $\ket{\psi(C^s_j)_{i+j}}$ is a state analogous to $\ket{\psi^\alpha_i}$ that
has $A^s$ matrices everywhere and $C^s_j$ inserted at site $(i+j)$.

\begin{figure}
\centering
\includegraphics[width=\columnwidth]{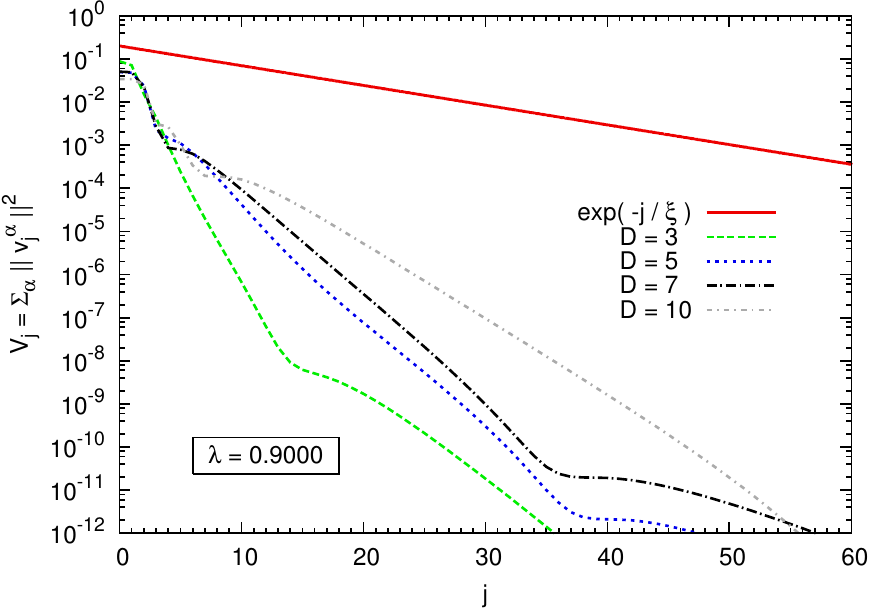}
\caption{(Color online) The quantity $V_j$ defined in Eq.\  \eqref{eq.def_Vj} for $\lambda = 0.9$ and various matrix dimensions $D$.
  The solid red line shows the function $0.2 \cdot \exp(-j/\xi)$ where $\xi$ is the exact correlation length from \eqref{eq.xi_TFIM}.
  $V_j$ displays exponential decay on a length scale that is always smaller than the correlation length $\xi$.}
\label{plot.Vj}
\end{figure}

To quantify the degree of localization of the excitations,
we study the squared norm of the vectors $\vec v_j$
\begin{align}
  V_j := \|\vec v_j\|^2 = \sum_\alpha |v_j^\alpha|^2 \ .
	\label{eq.def_Vj}
\end{align}
Figure \ref{plot.Vj} shows $V_j$ for various values of the bond dimension $D$ and
compares their dependence on $j$ to the decay of the correlation function $G_j$.
Clearly, the distributed (smeared out) contributions to the quasi-particle decay
exponentially with the distance $j$ from the center site.
\red{This agrees with the findings in Ref.\ \cite{haege13a}}.
With increasing matrix dimension the decay becomes slower and approaches the
decay of the correlation function. This is  is consistent with the finding
in Fig.\ \ref{plot.ZetaXi} illustrating that larger $D$ allows one to capture
longer correlations. For the numerics, it is very advantageous that the
decay of $V_j$ is always even faster than the decay of the correlation
function defined by the correlation length $\xi$, because this fact implies that the
representation can be truncated after a fairly small number of sites
$|j|\le j_\text{max}$.

\section{Spectral weight}
\label{sec:spectral_weight}

To illustrate the validity of the creation operator defined in \eqref{eq.def_AdQ} we compute the spectral weight $S^{xx}_\mathrm{1p}$.
For $\alpha = \beta = x$ Eq.\  \eqref{eq.Sxxq} becomes
\begin{align}
  S^{xx}_{\mathrm{1p}}(q) &= \bra{\psi_0} S_q^{x\dagger} a_q^\dagger \ket{\psi_0} \bra{\psi_0} a_q S_q^{x} \ket{\psi_0} \nonumber \\
  &=: |m_q|^2 \label{eq.Sxx_mq}
\end{align}
where $m_q$ is defined by $m_q = \bra{\psi_0} a_q S_q^{x} \ket{\psi_0}$.
Inserting the definition of $a_i^\dagger$ \eqref{eq.def_Ad_rs} and the Fourier transform of $\Sx_q$ we obtain
\begin{subequations}
\begin{align}
  m_q &= \frac{1}{L} \sum_{i,j,\alpha} v^{\alpha\ast}_q e^{iqr_j} e^{-iq r_i} \bra{\psi^\alpha_j} S^x_i \ket{\psi_0} \\
  &= \sum_{i,\alpha} v^{\alpha\ast}_q e^{iq r_i} \bra{\psi^\alpha_i} \Sx_0 \ket{\psi_0}
\end{align}
\end{subequations}
where the matrix elements $\bra{\psi^\alpha_i} \Sx_0 \ket{\psi_0}$ can be computed in analogy to the single-site operator in Eq.\  \eqref{eq.gse_per_site}.

\begin{figure}
\centering
\includegraphics[width=\columnwidth]{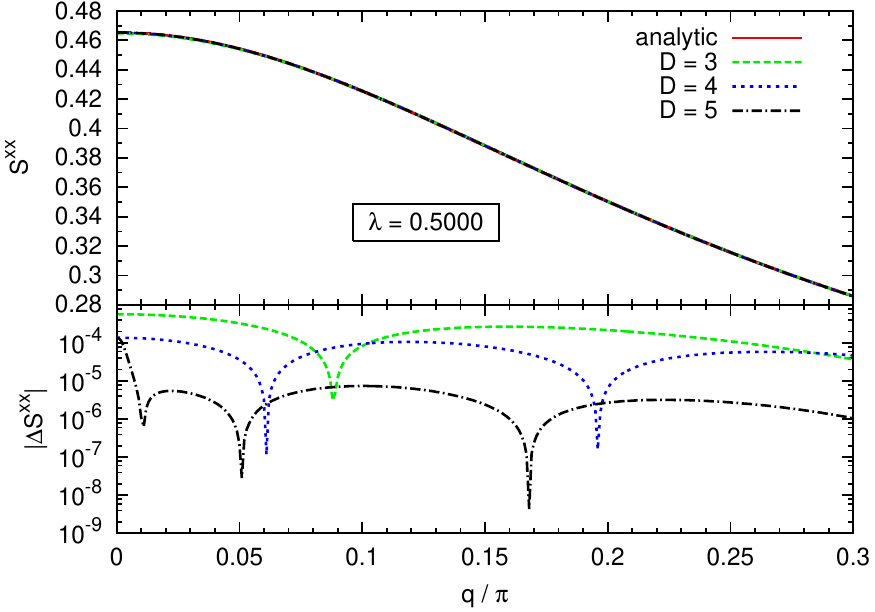}
\caption{(Color online) Upper panel: The spectral weight $S^{xx}_\mathrm{1p}(q)$ for $\lambda = 0.5$ and various matrix dimensions $D$.
Lower panel: The deviation of the iMPS results from Hamer's formula Eq.\
\eqref{eq.Sxx_Hamer}.
The plot interval $[0,0.3]$ is chosen to emphasize the deviation for small values of $q$
where $S^{xx}_\mathrm{1p}(q)$ has its maximum.}
\label{plot.Sxx1}
\end{figure}

\begin{figure}
\centering
\includegraphics[width=\columnwidth]{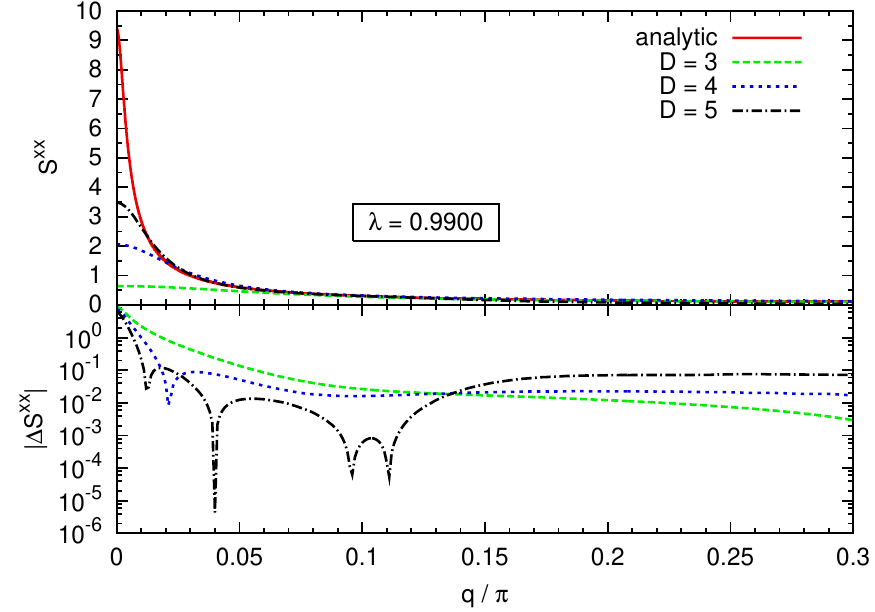}
\caption{(Color online) Upper panel: The spectral weight $S^{xx}_\mathrm{1p}(q)$ for $\lambda = 0.99$ and various matrix dimensions $D$.
Lower panel: The deviation of the iMPS results from Hamer's formula Eq.\  \eqref{eq.Sxx_Hamer}.
The plot interval $[0,0.3]$ is chosen to emphasize the deviation for small values of
$q$ where $S^{xx}_\mathrm{1p}(q)$ has its maximum.}
\label{plot.Sxx2}
\end{figure}

Figures \ref{plot.Sxx1} and \ref{plot.Sxx2} depict the spectral weight in comparison to the
analytical result Eq.\  \eqref{eq.Sxx_Hamer} for various values of $\lambda$ and $D$.
For smaller values of $\lambda$, see Fig.\ \ref{plot.Sxx1}, well away from the critical
point $\lambda=1$, the agreement is very good for all values of $q$.
Still, larger values of $D$
imply an even better agreement. For a value of $\lambda$ closer
to the critical point, the agreement is still good, see Fig.\ \ref{plot.Sxx2},
in view of the small values of $D$.
But in particular close to the almost diverging correlation at $q=0$, larger
values of $D$ are indispensable to capture the correct correlations.

\section{Conclusions}
\label{sec:outlook}

The objective of the present paper has been to sketch the roadmap to
a derivation of effective one-dimensional models by a numerical variational approach.
In particular, we have explicitly shown how the first step works, i.e.,
the systematic construction of a local creation operator acting on the ground
state. Thereby, bilinear terms in the Hamilton operator such as the dispersion
can be determined, cf.\ the generic Hamiltonian Eq.\ \eqref{eq.H_eff}.

We have shown how the matrix product state (MPS) formalism can be used to derive
effective models in terms of quasi-particles from microscopic local spin model Hamiltonians.
Based on transfer matrices, MPS work efficiently in the
thermodynamic limit (iMPS). Starting point of the MPS is the accurate determination of a
MPS representation of the ground state. This defines the vacuum of excitations
similar to the reference state in continuous unitary transformations \cite{knett03a}.

A side product of our ground state search algorithm are eigenmatrices with higher local energies.
We have shown how this side product can be exploited to
construct the elementary excited states. Such constructions
work very well for unique and for degenerate ground states.
In the latter case the elementary excitations generically are domain walls
between the degenerate ground states.

We derived an expression for the action of a
local quasi-particle creation operator on the ground state.
These quasi-particles are no longer completely local, but they are
found to be ``smeared out'' but
localized around one lattice site similar to Wannier states for
the band electrons. The approach is illustrated and tested
for the excitations in the transverse-field Ising model in one dimension
in the disordered strong-field phase as well as in the ordered Ising phase.
In the strong-field phase the elementary excitations are spin flips while
they are domain walls in the Ising phase.

It turns out that the quasi-particles are exponentially localized
on a length scale always smaller than the correlation length $\xi$.
In this way, the numerical representation of the elementary excitations
is well controlled. Using this definition, the one-particle contribution
to the spectral weight in the $xx$-channel has been computed.
The very good agreement with Hamer's formula \cite{hamer06b}
confirms his conjecture and strongly corroborates the validity of our approach.

What are the next steps on the roadmap to effective models from variational
approaches? In order to be able to determine the parts of the Hamiltonian
\eqref{eq.H_eff} which describe the decay of quasi-particles \eqref{eq.H_decay}
or the interaction of a pair of them \eqref{eq.H_int} we need to extend
the definition of single particle states to states with two-particles.
The key issues are a proper orthogonalization of states with excitations
at different sites. Furthermore, it must be ensured that
the two-particle state of two very distant quasi-particles equals the state obtained
from the successive application of the creation operator defined from
single-particle states. These issues are beyond the scope of the present
article, but represent future research. \red{The ultimate aim is
to be able to write down effective models in second quantization in terms
of the elementary excitations.}

\red{An interesting step towards this aim has been accomplished very recently
by the variational construction of scattering states of two elementary
excitations \cite{Haegeman2014}. But so far the explicit
construction of the effective model has not been realized.}

A longer-term vision consists in the generalization of the presented approach
to higher dimensions by passing from matrix product states to projected entangled pair states.
The conceptual issues and their solutions, for instance the construction
of Wannier type of local excitations, are the same in higher dimensions.
But the numerical handling is less efficient than in one dimension where the
thermodynamic limit is easily built-in by transfer matrices.

In summary, we are convinced that the construction of effective models
via numerical variational approaches constitutes an interesting and promising route
to capture the physics of strongly correlated systems.

\begin{acknowledgement}
We gratefully acknowledge the financial support of the Helmholtz virtual institute
``New States of Matter and Their Excitations''.
We also thank B.\ Fauseweh and N.A.\ Drescher for many helpful discussions.
\end{acknowledgement}

\appendix
\numberwithin{equation}{section}
\section{Ground state search algorithm}
\label{app:gs_search}

\subsection{Local fixed point iteration}

This appendix contains a more detailed discussion of the ground state search algorithm.
The starting point for a given guess for $A^s$ is computing $\mu_0$, $v$ and $u$.
For reasons of both computational efficiency and algorithmic simplicity, rescaling is done such that $\mu_0 = 1$ and $(u,v) = 1$.

Next, we consider the normalization constraint in Eq.\  \eqref{eq.min_Bs}, i.e., the division by the norm of the state $\ket{\psi(A^s,B^s)}$ that has $A^s$ matrices on all sites but one where the $B^s$ are inserted instead
\begin{align}
  \bracket{\psi(A^s,B^s)}{\psi(A^s,B^s)} = (u,\I^{(B,B)}[v]) \neq (u,v) \label{eq.norm_psi_B} \ .
\end{align}
The reference site where $A^s$ is replaced by  $B^s$ is labeled $i = 0$ which marks the center of the chain.
The norm in Eq.\  \eqref{eq.norm_psi_B} is a bilinear form in the coefficients of $B^s$, interpreted as a single $d\cdot D^2$-dimensional vector $\vec{B}$.
To see this, we recall how it is computed
\begin{subequations}
\label{eq.def_N}
\begin{align}
  (u,\I^{(B,B)}[v]) &= \Tr(u^\dagger \, \I^{(B,B)}[v]) \\
  &= \Tr\left( u^\dagger \, \sum_{s,s'} \delta_{ss'} B^{s'} v \, B^{s \dagger} \right)
	\\
  &= \sum_{\alpha'} \sum_{s,s'} \sum_{\beta,\beta',\alpha} u^\dagger_{\alpha'\beta} \delta_{ss'} B^{s'}_{\beta\beta'} v_{\beta'\alpha} (B^{s \dagger})_{\alpha\alpha'}
	\\
  &= \sum_{s,s'} \sum_{\alpha,\alpha',\beta,\beta'} B^{s \dagger}_{\alpha\alpha'} (\delta_{ss'} u^\dagger_{\alpha'\beta} v_{\beta'\alpha}) B^{s'}_{\beta\beta'} \ .
\end{align}
\end{subequations}
If one takes the three-tuples $(s,\alpha,\alpha') =: \tau$ and
$(s',\beta,\beta') = \tau'$ as single indices running from $1$ to $d \cdot D^2$,
the matrices $B^s$ correspond to a vector $\vec{B}$; we call this
step `vectorization'.
In this notation, the norm \eqref{eq.def_N} simply is a vector-matrix-vector product
\begin{subequations}
\begin{align}
  &\sum_{\tau,\tau'} \vec{B}^\dagger_{\tau} N_{\tau\tau'} \vec{B}_{\tau'} = \vec{B}^\dagger N \vec{B}
	\\
  \text{with } &N_{\tau\tau'} = N^{ss'}_{\alpha'\alpha,\beta'\beta} = \delta_{ss'} u^\dagger_{\alpha'\beta} v_{\beta'\alpha}
	\\
  \text{and }  &B_\tau = B_{\alpha\beta}^s \ .
\end{align}
\end{subequations}
If the vectorization $\vec{B}$ of $B^s$ is performed by concatenating all
columns of $B^s$ into a single column (cf.\ Eq.\  \eqref{eq.vec2mat}), a short calculation shows, that $N$ is given by the $d \cdot D^2 \times d \cdot D^2$ matrix
\begin{align}
\label{eq.def_N_Kronecker}
  N := \I_d \otimes v^T \otimes u^\dagger
\end{align}
where $v^T$ is the transpose of $v$.

In a next step we study  the application of the single-site operator $\Sz$ as it
occurs for instance in Eq.\  \eqref{eq.gse_TF}.
The algebraic structure is the same as in Eq.\  \eqref{eq.def_N},
but the identity operation is replaced by $\Sz$.
Thus, its expectation value with respect to $\ket{\psi(A^s,B^s)}$ is also a bilinear form in $\vec{B}$ given by
\begin{subequations}
\label{eq.def_Mi}
\begin{align}
  (u, S^{z\,(B,B)}[v]) &= \sum_{s,s'} \sum_{\alpha,\alpha',\beta,\beta'} B^{s \dagger}_{\alpha\alpha'} (\Sz_{ss'} u^\dagger_{\alpha'\beta} v_{\beta'\alpha}) B^{s'}_{\beta\beta'} \\
  &= \vec{B}^\dagger M^{[\Sz_0]} \vec{B} \\
  M^{[\Sz_0]} &:= \Sz \otimes v^T \otimes u^\dagger \label{eq.def_Mi_Sz}
\end{align}
\end{subequations}
where $\Sz$ is the local representation of the spin-$\frac{1}{2}$ $z$-operator
$S^z = \frac{1}{2} \sigma^z$, $\sigma^z$ being the $z$-Pauli matrix.

The Ising interaction in Eq.\  \eqref{eq.gse_Ising} is a bit trickier because
two sites are involved. Let them be the sites $i = 0$ and $i+1 = 1$.
Then the bilinear form reads
\begin{subequations}
\label{eq.def_Mi_mult_op}
\begin{align}
  (u, S^{x\,(B,B)}[ S^{x\,(A,A)}[v]]) &= \vec{B}^\dagger M^{[\Sx_0\Sx_1]} \vec{B} \\
  M^{[\Sx_0\Sx_1]} &:= \Sx \otimes \tilde v^T \otimes u^\dagger \label{eq.def_M_SxSx} \\
  \tilde v &:= S^{x\,(A,A)}[v] \label{eq.def_vtilde} \ .
\end{align}
\end{subequations}
One realizes in Eqs. \eqref{eq.def_Mi_Sz} and \eqref{eq.def_M_SxSx} that the structure of these expectation values is always the same.
It is the Kronecker product of the local operator matrix at site $i=0$ where the $B^s$ matrices are inserted and the Kronecker product of the boundary matrices $u$ $(\tilde u)$ and $v$ $(\tilde v)$.
If there is only the operator at site $0$ (Eq.\  \eqref{eq.def_Mi_Sz}), the
boundary matrices $u$ and $v$ are used without modifications.
If there are more operators involved, see Eq.\  \eqref{eq.def_Mi_mult_op}, the boundary
matrices are modified by applying these operators to them.
We denote this by the tilde symbol, see Eq.\  \eqref{eq.def_vtilde}.

Finally, the whole numerator of Eq.\  \eqref{eq.min_Bs} is given by the term
from Eq.\  \eqref{eq.def_Mi} and a sum of terms of the form \eqref{eq.def_Mi_mult_op}
\begin{subequations}
\begin{align}
  &\bra{\psi(A^s,B^s)_0} \left( \mathcal{H} - E_0 \right) \ket{\psi(A^s,B^s)_0}
	\\
  &= \sum_{i=-\infty}^{\infty} \bra{\psi(A^s,B^s)_0} \left( h_i - \frac{E_0}{L} \right) \ket{\psi(A^s,B^s)_0}
	\\
  &= \sum_{i=-\infty}^{-2} \vec B^\dagger \left( M^{[h_i]} - N\frac{E_0}{L} \right)
	\vec B
	\nonumber \\
  &+ \vec B^\dagger \left( M^{[\Sz_{-1}]} + M^{[\Sx_{-1} \Sx_{0}]} - N\frac{E_0}{L} \right) \vec B
	\nonumber \\
  &+ \vec B^\dagger \left( M^{[\Sz_0]} + M^{[\Sx_0 \Sx_1]} \right) \vec B
	\label{eq.BMB} \\
  &+ \sum_{i=1}^{\infty} \vec B^\dagger \left( M^{[h_i]} - N\frac{E_0}{L} \right)
	\vec B \\
  &= \vec B^\dagger M(\mathcal{H} - E_0) \vec B
\end{align}
\end{subequations}
where the matrices $M^{[h_i]}$ are given by
\begin{subequations}
\begin{align}
  M^{[h_i]} &= \begin{cases}
		\I_d \otimes \tilde v^T \otimes u^\dagger & \text{if } i > 0
		\\
		\I_d \otimes v^T \otimes \tilde u^\dagger & \text{if } i < 0
              \end{cases} \\
  \text{with } \tilde v &= T^{i-1}[h_i[v]] \\
  \text{and } \tilde u &= T^{\dagger\, i-2}[h_i^\dagger[u]]
\end{align}
\end{subequations}
and $N$ is the matrix defined in Eq.\  \eqref{eq.def_N_Kronecker}.

Note, that in Eq.\  \eqref{eq.BMB} one summand $-N\frac{E_0}{L}$ is omitted.
The lowest eigenvalue $\mu_0$ then converges to the variational ground state energy per lattice site rather then to zero.
As a rule of thumb in numerics it is better to search for finite values than zero, especially when dealing with relative errors.

As shown in Eq.\  \eqref{eq.Hj_decay}, the boundary contributions quick{-}ly converge to zero for $|i| \gg 0$ so that the sums can be truncated after a finite number of sites, i.e., comprising a finite tractable number of terms.

Eq.\  \eqref{eq.min_Bs} can be recast into the form
\begin{align}
  \vec{B}^\dagger M(\mathcal{H},A^s) \vec{B} - \frac{E_0}{L} \vec{B}^\dagger N(A^s)
	\vec{B} = 0\ .
\end{align}
The minimization of $E_0$ amounts up to finding roots of the derivative with respect to $\vec{B}^\dagger$. This yields the generalized EVP in Eq.\  \eqref{eq.GSE_EVP}.

The complete algorithm runs as follows:
\begin{itemize}
  \item[1.] Start with an initial guess for $A^s$.
  \item[2.] Generate the matrices $M$ and $N$.
  \item[3.] Solve the generalized EVP Eq.\  \eqref{eq.GSE_EVP}.
  \item[4.] Break if $B_0^s = A^s$ within a given tolerance.\\
    A local minimum of the ground state energy is a fixed point of the iteration which satisfies this condition.
  \item[5.] The eigenvector $\vec{B}_0$ with lowest energy $\epsilon_0$ is chosen as new guess for $A^s$.
    Go to step $2$.
\end{itemize}
If no better initial guess is available, start with a random matrix set in step $1$.
In cases where multiple values of a system parameter are to be investigated, the converged solution for a nearby value generically constitues a good initial guess.
Since there is no way to determine if a solution is also a global minimum, the algorithm is terminated if a fixed point $B_0^s = A^s$ is reached within numerical tolerance.

\red{In contrast, conventional MPS-based iDMRG, see for instance Ref.\
\cite{mccull08}) follows the spirit of White's original DMRG for infinite systems \cite{white92}.
In each iteration, the system is incremented by adding one unit cell in the center of the chain, thereby increasing the bond dimension locally
which is then truncated to its original value using the most relevant
part of the density matrix. Convergence is reached when the matrices $A^{s}_{n+1}$
obtained for the added unit cell are the same as the $A^{s}_{n}$ obtained in the previous iteration within some preset tolerance.}

\subsection{Fine tuning}
\label{ss:fine-tuning}

Figure \ref{plot.ZetaXi} shows that close to criticality the results improve visibly  with growing matrix dimension $D$.
This is so because a larger bond dimension allows more correlations to be represented.
At or close to the strong-field limit, however, a large bond dimension may actually be disadvantageous due to a certain lack of entanglement.
The matrices $A^s$ are too large to encode the small amount of entanglement
in the system. This makes itself felt in the norm matrix $N$ in Eq.\  \eqref{eq.GSE_EVP}
 becoming singular or having very small eigenvalues in magnitude.
Solving the generalized EVP involves division by the eigenvalues of $N$
so that the generalized EVP is ill-defined if $N$ is singular or close to it.

But Eq.\  \eqref{eq.def_N_Kronecker} shows that in the gauge where $v = \I_D$, $N$ is diagonal and holds $d\cdot D$ copies of $u$.
This matrix $u$ is the reduced density matrix of the left subsystem
if the system is split into a left  and a right part which is traced out, as is done in DMRG.
Therefore, omitting the vectors corresponding to small eigenvalues of $u$ is a systematically controlled way to focus on the relevant subspace.
The null space of $N$ is projected out which also avoids numerical instabilities,
 cf.\ Eq.\  \eqref{eq.gen_to_std_EVP}.

Let $D'$ be the dimension kept in the truncated density matrix $u$.
Projecting out the null space of $N$ results in $d\cdot D \cdot D'$ eigenvectors of
Eq.\  \eqref{eq.GSE_EVP} instead of $d\cdot D^2$.
This is also efficient in the subsequent calculations by
speeding up the dispersion calculation since the initial dimension of $H_q$ and $N_q$ is reduced to $d\cdot D \cdot D'$.

A last aspect in the ground state optimization concerns the iteration
$B_0^s\to A^s$. As mentioned in the main text, it is not at all clear
whether taking the matrix set $B_0^s$ found for a single site at all sites
indeed improves the ground state. But we can ensure that the
variational ground state energy is reduced in each iterative step by
performing a linear search using the ansatz
\begin{subequations}
\begin{align}
  E_0^\ast &= \min_{x \in (0,\pi/2)} E( \cos(x) A^s + \sin(x) B_0^s )
	\\
  \rightarrow B_0^{s\ast} &= \cos(x_\mathrm{min}) A^s + \sin(x_\mathrm{min}) B_0^s
\end{align}
\end{subequations}
which interpolates between $A^s$ and $B_0^s$. The one-dimensional minimization of $E(x)$
as function of $x$ is numerically robust.

As can be seen from Eq.\  \eqref{eq.gse_per_site}, the ground state energy is a highly nonlinear function of the coefficients of $A^s$.
Any multi-dimensional minimizer, that does not rely on derivatives, can be used to find
 a minimum, e.g., the method of conjugate directions or simulated annealing.
On the one hand, our experience shows that these routines converge at a much slower rate than the algorithm described above.
On the other hand, however, the fixed point iteration may fail to converge if the initial guess is too far away from an optimal solution.
Therefore, we actually use a hybrid algorithm.
First, a couple of iterations of simulated annealing are preformed, yielding a good inital guess. Then this guess is used for the above given algorithm
which breaks if $B_0^{s\ast}=A^s$ within numerical tolerance.

\section{Creation operator}
\label{app:creation_op}

This short appendix contains general technical details of the computation of the representation of the creation operators $a_q^\dagger$ and $a_i^\dagger$.

The generalized EVP Eq.\  \eqref{eq.dispersion_EVP} is solved for each momentum value
$q$ independently. Therefore, an arbitrary phase may always occur between the eigenvectors $\vec{v}_q$ and $\vec{v}_{q+\Delta q}$ where $\Delta q$ is the sampling interval in $q$-space.
In order for the Fourier transform Eq.\ \eqref{eq.def_vj} back to real space
to yield well-localized components $v_j^\alpha$ each component $\vec{v}_q^\alpha$ must be a smooth, $2\pi$-periodic function in $q$.

In order to ensure this smoothness we employ a two step process.
First, we fix the phase between adjacent vectors only separated by $\Delta q$
to zero by setting
\begin{align}
  \vec{v}_q = \frac{\vec{v}_q}{\Phi_q} \qquad \text{with }
	\Phi_q := \frac{\vec{v}_{q-\Delta q}^\dagger
	\vec{v}_q}{\|\vec{v}_{q-\Delta q}\| \|\vec{v}_q\|} \ .
\end{align}
But this choice of vanishing phase is still somewhat arbitrary.
More generally, a phase of the order of $\Delta q$ could occur between adjacent vectors.

In practice, we check for the phase between $\vec{v}_{q=-\pi}$ and $\vec{v}_{q=\pi}$ after the above smoothing process. It should vanish due to $2\pi$-periodicity, but
this may not be the case. To restore $2\pi$-periodicity in a second step, the accumulated phase between $\vec{v}_{q=-\pi}$ and $\vec{v}_{q=\pi}$ is distributed evenly over the whole Brillouin zone. This procedure
results in the very fast decaying quasi-particle representation
presented in Fig.\ \ref{plot.Vj}.

\section{Ground state degeneracy}
\label{app:gs_degeneracy}

As mentioned above, ground state degeneracy is reflected in the spectrum of the transfer matrix $T$.
If an exact iMPS prepresentation of the ground state exists at finite $D$, this results in a degeneracy of the largest absolute value of the eigenvalue $\mu_0$.
This is for instance the case for the Majumdar-Ghosh model \cite{majum69a,majum69b,majum69c} that has an exact ground state iMPS representation at $D = 3$.
If no exact iMPS exists for finite $D$, as is the case for the  TFIM, one still observes that
\begin{align}
  \frac{|\mu_0|}{|\mu_1|} \to 1 \qquad \text{for} \qquad D \to \infty \ ,
\end{align}
i.e., there is an asymptotic degeneracy.

However, in some cases analytical considerations may help.
For the TFIM in the Ising phase we know that the ground state $\ket{\psi^+_0}$
with magnetization in positive $S^x$ direction can be transformed into
the ground state $\ket{\psi^-_0}$ with magnetization in negative $S^x$ direction
by a $\pi$-rotation about the $z$-axis
\begin{align}
  \ket{\psi^-_0} = e^{i \pi S^z} \ket{\psi^+_0} \ .
	\label{eq.TFIM_gs_trafo}
\end{align}
This rotation is a non-local operation, but it is the same for all sites
\begin{align}
  e^{i\pi S^z} \ket{\psi^+_0} &=
  \Tr\left[ \left( \sum_{s_1,s_1'} (e^{i\pi S^z})_{s_1 s_1'} A^{s_1'} \right) \times \cdots  \right.
	\nonumber \\
  &\qquad\quad\left. \times \left( \sum_{s_L,s_L'} (e^{i\pi S^z})_{s_L s_L'} A^{s_L'} \right) \right] \ket{\{s_i\}} \ .
\end{align}

Recall that $e^{i\pi S^z} = \cos(\pi/2) \I + i\sin(\pi/2) \sigma^z = i\sigma^z$ since $S^z = \frac{1}{2}\sigma^z$ where $\sigma^z$ is the $z$-Pauli matrix and
$\sigma^{z\,2} = \I$. The phase factor $i$ is a special case of the gauge transformation Eq.\  \eqref{eq.gauge} and can be dropped.
In the iMPS representation the spin rotation results in a relative sign between the two matrices
\begin{align}
  \ket{\psi^+_0} : \{ A^1, A^2 \} \quad \to \quad \ket{\psi^-_0} : \{ A^1, -A^2 \} \ .
\end{align}
Note, that this representation of $\ket{\psi^-_0}$ is not canonical anymore.
But it can be made canonical by the algorithm mentioned in the main text and
presented  in Ref.\ \cite{orus14}. The resulting density matrix $u$ turns out
to the be same as for $\ket{\psi^+_0}$.

The two ground states can be distinguished by the sign of the magnetization in $S^x$-direction
\begin{align}
  M_x = \bra{\psi^\pm_0} S^x \ket{\psi^\pm_0}
\end{align}
which serves as the order parameter in the Ising regime.
The ground state search algorithm produces either one or the other realization, not a superposition of both.
This is due to the fact, that an iMPS representation strongly favors \red{pure and finitely correlated states}.
Obviously, this is the case for either state $\ket{\psi^\pm_0}$ but not for their superposition.
Near the degeneracy of $\mu_0$ one eigenvalue dominates numerically and the algorithm
converges to the corresponding eigenvector as fixed point.
Which state will finally be selected depends on the initial guess for $A^s$.

For general models, however, it may not be clear, how degenerate ground states are connected, i.e.,
if there is an analytically applicable transformation such as Eq.\ \eqref{eq.TFIM_gs_trafo}.
But the occurrence of degenerate transfer matrices is a strong indicator for degenerate ground states.
The comparison of the expectation values of possible order parameters for different ground state solutions may help to distinguish them.

\subsection{Dispersion calculation with degenerate ground state}

As mentioned in Sect. \ref{sec:model}, the elementary excitations in the strong-field and in the Ising phase are qualitativly different.
In the strong-field regime, they consist of a local perturbation of the otherwise uniform ground state as described by the ansatz in Eq.\  \eqref{eq.def_Psi_ARS}.
In the Ising phase, the elementary excitations are domain walls separating regions of different ground state, i.e., the excitations  are non-local.
The domain wall character requires a modified ansatz
\begin{align}
  \ket{\psi^\alpha_i} = \sum_{\{s_i\}} \Tr(A^{s_1} \cdots A^{s_{i-1}} B^{s_i}_\alpha \tilde A^{s_{i+1}} \cdots \tilde A^{s_L}) \ket{\{s_i\}}
	\label{eq.def_Psi_ARS_deg}
\end{align}
where $\tilde A^s$ describes an alternative ground state.
To obtain the appropriate eigenmatrices $\vec B_{\alpha}$ a different kind of generalized EVP has to be solved once the two ground states are known
\begin{align}
  \bar M(A^s,\tilde A^s,\mathcal{H}) \vec B = \epsilon \bar N(A^s,\tilde A^s) \vec B
\end{align}
where the matrices $\bar M$ and $\bar N$ are built in the same way as $M$ and $N$ from Eq.\  \eqref{eq.GSE_EVP}, see Appendix \ref{app:gs_search} for details.
But there is one important difference.
Instead of the eigenmatrix $v$ of $T = \sum_s A^{s\ast} \otimes A^s$ the eigenmatrix $\tilde v$ of $\tilde T = \sum_s \tilde A^{s\ast} \otimes \tilde A^s$ has to be used.
In this case, the eigenmatrix $\vec B_{\alpha=0}$ no longer is one of the ground state matrix sets $A^s$ or $\tilde A^s$.

For domain wall excitations it is not directly evident, that the $\ket{\psi_i^\alpha}$ and their Fourier transforms $\ket{\psi_q^\alpha}$ are orthogonal to the ground states although this still holds.
This is simply due to the fact that in the thermodynamic limit different ground states are orthogonal.
Therefore, any excited state that contains a domain wall and thus regions of both ground states is orthogonal to $\ket{\psi^\pm}$ for the infinite system.

To see this in the iMPS representation we consider the overlap of the two ground states
\begin{align}
  \bracket{\psi^-}{\psi^+} = \bar u^\dagger \bar \mu_0^{L} \bar v
	\label{eq.overlap}
\end{align}
where $\bar \mu_0$ is the largest eigenvalue in absolute value
and $\bar u$, $\bar v$ are the corresponding left and right eigenvectors of
$\bar T = \sum_s \tilde A^{s\ast} \otimes A^s$.
Since $\ket{\psi^+} \neq \ket{\psi^-}$, $\bar\mu_0 \neq \mu_0$ holds.
It turns out, that $|\bar \mu_0| < 1$.
Therefore, the overlap \eqref{eq.overlap} tends to zero as $L \to \infty$.
This implies that the ground states are orthogonal in the thermodynamic limit.

Thus, the dispersion calculation is conceptually the same for degenerate ground states as for the non-degenerate case.
Only the dimension of the matrices $H_q$ and $N_q$ in Eq.\  \eqref{eq.dispersion_EVP} may be increased by one because $\vec B_{\alpha=0}$ may also represent an excitation.
In the computation of the matrices $H_q$ and $N_q$ one has to account for the different ground states.
For instance, the overlap of two states as defined in Eq.\  \eqref{eq.def_Psi_ARS_deg} is given by ($j < 0$)
\begin{align}
  \bracket{\psi_j^\alpha}{\psi_0^\beta} &= (u, \I^{(B_\alpha,A)}[ \bar T^{j-1}[ \I^{(\tilde A,B_\beta)}[\bar v]]]) \ .
\end{align}


\end{document}